\documentclass[12pt, draftclsnofoot, onecolumn]{IEEEtran}
\usepackage{amsfonts,amssymb}
\usepackage{amsmath}
\usepackage{array}
\usepackage{cite}
\usepackage{url}
\usepackage{xcolor}
\usepackage{cite,graphicx,amsmath,amssymb}
\usepackage{subfigure}
\usepackage{fancyhdr}
\usepackage{mdwmath}
\usepackage{mdwtab}
\usepackage{amsthm}
\usepackage{booktabs}
\usepackage{multirow}
\usepackage{longtable}
\usepackage{rotating}

\newtheorem{remark}{Remark}
\newtheorem{theorem}{Theorem}
\newtheorem{definition}{Definition}
\newtheorem{approximation}{Approximation}

\newtheorem{lemma}{Lemma}

\newtheorem{corollary}{Corollary}

\newtheorem{proposition}{Proposition}

\usepackage[breaklinks,colorlinks,linkcolor=black,citecolor=black,urlcolor=black]{hyperref}

\begin{document}

\title{Is the Envelope Beneficial to Non-Orthogonal Multiple Access?}

\author{Ziyi~Xie,~\IEEEmembership{Student Member,~IEEE,} Wenqiang~Yi,~\IEEEmembership{Member,~IEEE,} Xuanli~Wu,~\IEEEmembership{Member,~IEEE,} Yuanwei~Liu,~\IEEEmembership{Senior Member,~IEEE,} and Arumugam~Nallanathan,~\IEEEmembership{Fellow,~IEEE}
	
\thanks{Z. Xie and X. Wu are with the School of Electronics and Information Engineering, Harbin Institute of Technology, Harbin 150001, China (email: \{ziyi.xie, xlwu2002\}@hit.edu.cn).}

\thanks{W. Yi, Y. Liu, and A. Nallanathan are with the School of Electronic Engineering and Computer Science, Queen Mary University of London, E1 4NS, U.K. (email: \{w.yi, yuanwei.liu, a.nallanathan\}@qmul.ac.uk).}

\thanks{Part of this work was submitted to IEEE International Conference on Communications (ICC), 2023 \cite{0conference}.}
 
 }



\maketitle

\vspace{-0.3 cm}
\begin{abstract}
Non-orthogonal multiple access (NOMA) is capable of serving different numbers of users in the same time-frequency resource element, and this feature can be leveraged to carry additional information. 
In the orthogonal frequency division multiplexing (OFDM) system, we propose a novel enhanced NOMA scheme, called NOMA with informative envelope (NOMA-IE), to explore the flexibility of the envelope of NOMA signals. In this scheme, data bits are conveyed by the quantified signal envelope in addition to classic signal constellations. The subcarrier activation patterns of different users are jointly decided by the envelope former. At the receiver, successive interference cancellation (SIC) is employed, and we also introduce the envelope detection coefficient to eliminate the error floor. Theoretical expressions of spectral efficiency and energy efficiency are provided for the NOMA-IE. Then, considering the binary phase shift keying modulation, we derive the asymptotic bit error rate for the two-subcarrier OFDM subblock. Afterwards, the expressions are extended to the four-subcarrier case. The analytical results reveal that the imperfect SIC and the index error are the main factors degrading the error performance. The numerical results demonstrate the superiority of the NOMA-IE over the OFDM and OFDM-NOMA, especially in the high signal-to-noise ratio (SNR) regime.
\end{abstract}

\vspace{-0.2 cm}
\begin{IEEEkeywords}
Bit error rate, envelope, index modulation, non-orthogonal multiple access, OFDM, successive interference cancellation.
\end{IEEEkeywords}






\vspace{-0.2 cm}
\section{Introduction} 
Due to emerging Internet-enabled applications, such as extended reality (XR) services and telemedicine, the next-generation wireless networks require a more robust communication quality~\cite{6G}.
Non-orthogonal multiple access (NOMA) has been regarded as a promising multiple access candidate for future networks~\cite{NOMA1,NOMA2,NOMA3}. 
Different from the orthogonal multiple access (OMA), multiple users in NOMA can be served by the same time-frequency resource element by utilizing the code or power domain. Particularly in the power-domain NOMA, the superposition coding is applied at the transmitter for power multiplexing, and the successive interference cancellation (SIC) is adopted at the receiver for detection. When considering a perfect SIC and user fairness, it has been theoretically proved that the NOMA\footnote{In this work, NOMA refers to the power-domain NOMA.} has a larger sum rate than OMA in high signal-to-noise ratio (SNR) regimes~\cite{DingPairu}. Recent contributions focused on the Shannon channel capacity evaluation of NOMA in various scenarios\cite{NOMAUAV,NOMAout2,NOMAout3,NOMAcapa}. With the error-free decoding assumption, it is difficult for the initial NOMA studies to evaluate the communication quality under practical modulation schemes, especially with a non-ideal SIC process.

Another interesting property of NOMA in multi-carrier systems is that the number of users in the same subcarrier is more than one. This quantified signal envelope can be utilized to carry additional information. For example, in the two-user case, all possible numbers of users in a subcarrier are 0, 1, and 2. Therefore, the envelope is able to convey at most $\log_23$ bits of information. A similar concept can be found in an OMA scheme called orthogonal frequency division multiplexing with index modulation (OFDM-IM)~\cite{IMOFDMBasar, IM1, IM2}. OFDM-IM is developed from the orthogonal frequency division multiplexing (OFDM), which has been involved in wireless communication standards since Long Term Evolution (LTE) due to its capability of combating the intersymbol interference caused by the frequency-selective fading channel.
In OFDM-IM, if there is no signal in a subcarrier, the subcarrier is inactive; otherwise, it is active.
Exploiting the subcarrier orthogonality in conventional OFDM, OFDM-IM introduces a new subcarrier-index dimension to carry information in addition to the classical signal constellations, i.e., those binary subcarrier activation patterns convey additional information~\cite{IMSNR, IMsurvey, BasarSurv}. OFDM-IM has been regarded as an effective scheme providing either bit error rate (BER) performance enhancement or power-efficiency improvement~\cite{SubIM}. To explore the potential of the OFDM-IM in multi-user cases, several works have investigated the combination of the OFDM-IM framework and NOMA~\cite{IMNOMA, IMNOMAopen}. However, these works simply applied the OFDM-IM waveform to different NOMA users, and the performance enhancement is limited in specific conditions. In this work, we propose a novel {\it NOMA with informative envelope (NOMA-IE)} scheme, where the activation patterns of different users are jointly decided by the transmitter.
 
\vspace{-0.3 cm}
\subsection{Related Works}
\vspace{-0.2 cm}
In practical implementation, when considering the finite constellation input, the decoding error cannot be ignored, especially during the SIC procedure\cite{IETBER}. Therefore, recent research has paid attention to the error performance of NOMA. The exact closed-form bit error rate (BER) under fading channels was firstly derived in \cite{IETBER}, where the binary phase shift keying (BPSK) and quadrature phase shift keying (QPSK) are selected for the far user and the near user, respectively. In \cite{SER}, the authors provided the exact symbol error rate with arbitrary ordered pulse-amplitude modulation (PAM) and quadrature amplitude modulation (QAM). Nevertheless, these works are limited to the two-user scenario. The authors in \cite{BERBPSK} considered the BPSK modulation and investigated the BER performance when the number of users is arbitrary. A NOMA real-time test system was employed to validate the results. Both the asymptotic and exact BER expressions of the QAM with arbitrary modulation order were obtained in multi-user downlink NOMA systems\cite{BERArbi}.

Some simple combinations of the NOMA concept and OFDM-IM have been studied in previous works. In \cite{HybridIM}, the authors focused on a hybrid NOMA scheme exploring power domain multiplexing of OFDM and OFDM-IM. The proposed scheme has better error performance and achievable rate than the conventional OFDM-based NOMA.
In \cite{IMNOMA} and \cite{IMNOMAopen}, all NOMA users adopt the OFDM-IM for downlink transmissions, and the scheme is termed IM-NOMA. Different user requirements can be fulfilled by adjusting power allocation factors and subcarrier activation ratios.
In \cite{IMNOMAup}, the uplink NOMA was considered, and the OFDM-IM was proved to reduce the impact of the collision on the ultrareliable low-latency communication. In these works, data bits of NOMA users are mapped to the subcarrier indices respectively before the superposition coding, and the envelopes of different signals are independent. In addition, at the receiver, the SIC is employed as in NOMA. However, the current integration of NOMA and OFDM-IM brings problems in terms of theoretical performance analysis and practical realization. For tractability, the previous works derived the approximated BER expressions by ignoring the interference caused by NOMA. Moreover, the SIC results in the error floor in the OFDM-IM framework, which degrades the error performance when NOMA users have close power levels\cite{efloor}. 

\vspace{-0.4 cm}
\subsection{Motivations and Contributions}
\vspace{-0.1 cm}
As we have discussed, although the OFDM-IM framework has the potential to leverage the signal envelope as a new degree of freedom to boost the performance of NOMA, multiple problems emerge when simply combining two techniques. In this work, considering practical modulation schemes, a novel NOMA-IE design is proposed. In this scheme, we provide a better solution to integrate NOMA with the OFDM-IM framework. More specifically, we aim at answering the following questions:
\begin{itemize}
	\item {\bf Question 1}: How to derive more accurate BER expressions when considering the intra-cluster interference in NOMA?
	\item {\bf Question 2}: How to reduce or eliminate the error floor caused by SIC in NOMA-IE?
	\item {\bf Question 3}: With the additional flexibility from the envelope, how does the performance superiority of OFDM-IE over conventional OFDM or OFDM-NOMA behave?
\end{itemize}

We provide a detailed analysis and discussion of the proposed NOMA-IE. The main contributions of this paper are summarized as follows:
\begin{itemize}
	\item Considering a two-user case, we propose a novel NOMA-IE scheme based on the OFDM-IM framework in downlink transmissions. At the transmitter, an envelope former module is employed to jointly decide the subcarrier activation patterns of two NOMA users. At the receiver, we adopt the SIC decoding as in conventional NOMA. The envelope detection coefficient is introduced to the detection procedure for the signal of the far user.
	\item We provide accurate expressions of spectral efficiency and energy efficiency. Then we mainly analyze the theoretical BER performance of the NOMA-IE. With the consideration of intra-cluster interference, we derive the asymptotic BER expressions for two NOMA users in the two-subcarrier OFDM subblock to answer {\bf Question 1}. These expressions are quite accurate in the high SNR regime. The results can be extended to arbitrary multi-subcarrier subblocks, and we take the four-subcarrier case as an example in this work. The analytical results show that the imperfect SIC and the index error significantly degrade the error performance.
	\item We obtain that the value of the power allocation coefficient affects the error floor when employing the detection method in previous works. Since the design of NOMA-IE avoids the constellation overlap in NOMA signals, the error floor in NOMA-IE can be eliminated by adjusting the envelope detection coefficient, and {\bf Question 2} is answered. We also show that an appropriate envelope detection coefficient value helps to achieve a low BER.
	\item We answer {\bf Question 3} by the numerical results. The numerical results validate our analysis and demonstrate that: 1) compared with the conventional OFDM, the OFDM-IM framework reduces the BER at high SNR; 2) NOMA-IE with the feasible envelope detection coefficient has better error performance than the IM-NOMA in most cases; 3) under equal spectral efficiency and energy efficiency, the proposed NOMA-IE outperforms OFDM-NOMA in the high SNR regime and outperforms the conventional OFDM with the most values of power allocation coefficient. 
\end{itemize}

The rest of the paper can be summarized as follows. In Section II, the system model of NOMA-IE is presented. In Section III, we provide the performance metrics considered in this paper as well as a preliminary analysis of NOMA-IE. The theoretical error performance of NOMA-IE is investigated in Section IV. In Section V, numerical results are illustrated. Finally, Section VI concludes the paper.

{\it Notation:} $(\cdot)^T$ denotes the transpose operation. $|\cdot|$ denotes the absolute value if applied to a complex number or the cardinality if applied to a set. $\left\| \cdot \right\|$ denotes the Frobenius norm. $\binom{n}{k} = \frac{n!}{k!(n-k)!}$ denotes the binomial coefficient and $\left\lfloor \cdot \right\rfloor$ is the floor function. $Q(\cdot)$ denotes the Gaussian Q-function. The probability of an event is denoted by ${\rm Pr}(\cdot)$. $\det({\bf A})$ stands for the rank of ${\bf A}$. ${\bf I}_N$ is the identity matrix with the dimension $N \times N$.  For the vector ${\bf x} = [x(1),x(2),...,x(n)]^T$, we define ${\bf X} = {\rm diag}\{x(1),x(2),...,x(n)\}$. $X \sim {\cal CN}(0,\sigma^2)$ represents the distribution of a circularly symmetric complex Gaussian random variable $X$ with variance $\sigma^2$.

\section{System Model}
\begin{figure*}[t!] 
	\centering
	\subfigure[]{\includegraphics[width=5.7in]{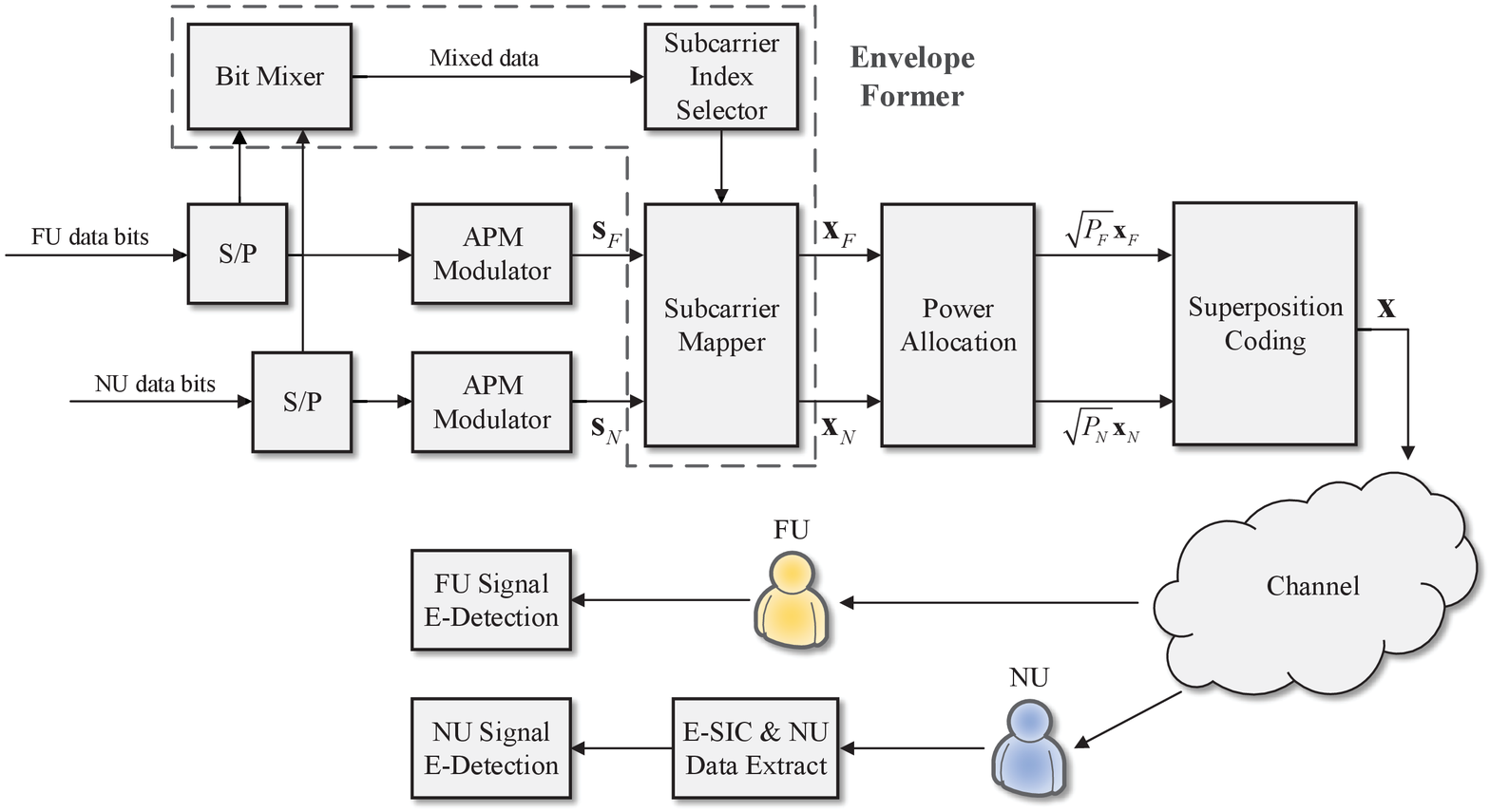}
		\label{sys: fig_a}}
	\hfil
	\subfigure[]{\includegraphics[width=3.2in]{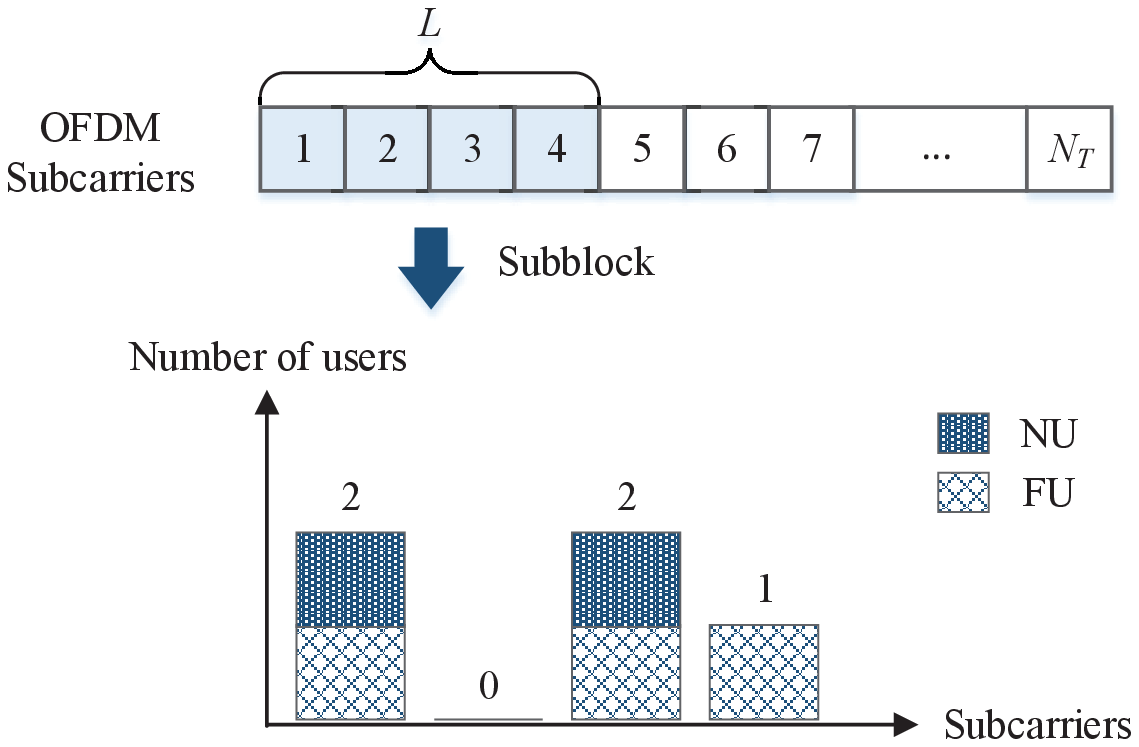}
		\label{sys: fig_b}}
	\caption{Illustration of system model: (a) block diagram of the proposed NOMA-IE scheme; (b) a mapping example when $L=4$, $K_F = 3$, $K_N = 2$, and mixed index bits are $\left[0,1,1 \right]$.
	}
	\label{fig1}
\end{figure*}

As shown in Fig. \ref{fig1}, we consider a downlink NOMA-IE scheme based on the OFDM-IM framework\cite{BasarSurv}, where a near user (NU) and a far user (FU) share total $N_T$ subcarriers. This system operates in a Rayleigh fading environment, and the fading gains on different subcarriers are independent and identically distributed. 

\vspace{-0.2 cm}
\subsection{Transmitter Design with Envelope Forming}
Similar to the conventional OFDM-IM, the total $N_T$ subcarriers are divided into $G = N_T/L$ subblocks, each of which consists of $L$ subcarriers as shown in Fig. \ref{sys: fig_b}. Thus data bits of both users are equally split into $G$ groups for transmission. In each data group, part of the data bits called index bits is for the decision on the active subcarrier indices while the other part called symbol bits is modulated on active subcarriers. For subblock $g \in \{1,2,...,G\}$, we use ${\bf I}_{g,u}$ to denote the indices set of active subcarriers, where $u \in \{N, F\}$. 
\begin{definition}\label{def: IC-NOMA}
(Principle of Envelope Forming) In the NOMA-IE scheme, the set of active subcarriers for the NU is a subset of that for the FU, i.e., ${\bf I}_{g,N} \subseteq {\bf I}_{g,F}$\footnote{It can be easily extended to $N$-user NOMA clusters ($N>2$). We will illustrate the effectiveness of the definition later in this paper. }.
\end{definition}

The number of active subcarriers $K_{g,u}$ in the subblock $g$ is the cardinality of ${\bf I}_{g,u}$, i.e., $K_{g,u} = \left| {\bf I}_{g,u} \right|$, and hence $K_{g,N} \le K_{g,F} \le L$ holds. 
\begin{remark}
The special cases of the NOMA-IE are other classic OFDM-based schemes. When $K_{g,N} = 0$ and $K_{g,F} = L$, the NOMA-IE becomes the conventional OFDM. When $K_{g,N} = 0$ and $0 <K_{g,F} < L$, the NOMA-IE is the OFDM-IM. When $K_{g,N} =K_{g,F} = L$, it can be regarded as the OFDM-NOMA \cite{OFDMNOMA}.
\end{remark}

Since data bits conveyed by different subblocks are independent, we focus on one subblock in the rest of the paper, and we denote $K_{g,u} = K_u$, where $K_u$ is a constant. Considering the $M_u$-ary amplitude and phase modulation (APM) scheme, a total of
\begin{align}
	{m}_{F} &= \underbrace{\left\lfloor {\log_2 \binom{L}{K_{F}}} \right\rfloor}_{\mbox {Index bits}} + \underbrace{K_{F}\log_2M_{F}} _{\mbox {Symbol bits}},
\end{align}
bits are conveyed by the FU signal per subblock. In each subblock of the NU signal, total data bits are
\begin{align}
	{m}_{N} &= \underbrace{\left\lfloor {\log_2 \binom{K_{F}}{K_{N}}} \right\rfloor}_{\mbox {Index bits}} + \underbrace{K_{N}\log_2M_{N}}_{\mbox {Symbol bits}}.
\end{align}

\begin{table*}
	\centering
	\caption{An Example of the NOMA-IE Mapping Table for $L=4$, $K_{F}=3$, and $K_{N}=2$} \label{table: mapping}	
	\begin{tabular}{
			m{2.4cm}<{\centering} |m{1.9cm}<{\centering}
			|m{3.4cm}<{\centering} |m{1.6cm}<{\centering} 
			|m{3cm}<{\centering}}
		\hline \hline
		Mixed index bits & ${\bf I}_{F}$ & FU subblock ${\bf x}_F$ & ${\bf I}_{N}$ & NU subblock ${\bf x}_N$  \\ \hline
		$[0,0,0]$ & $\{2,3,4\}$ & $[0,s_{F}(1),s_{F}(2),s_{F}(3)]^T$ & $\{3,4\}$ & $[0,0,s_{N}(1),s_{N}(2)]^T$ \\ 
		$[0,0,1]$ & $\{2,3,4\}$ & $[0,s_{F}(1),s_{F}(2),s_{F}(3)]^T$ & $\{2,3\}$ & $[0,s_{N}(1),s_{N}(2),0]^T$ \\ 
		$[0,1,0]$ & $\{1,3,4\}$ & $[s_{F}(1),0,s_{F}(2),s_{F}(3)]^T$ & $\{3,4\}$ & $[0,0,s_{N}(1),s_{N}(2)]^T$ \\
		$[0,1,1]$ & $\{1,3,4\}$ & $[s_{F}(1),0,s_{F}(2),s_{F}(3)]^T$ & $\{1,3\}$ & $[s_{N}(1),0,s_{N}(2),0]^T$ \\
		$[1,0,0]$ & $\{1,2,4\}$ & $[s_{F}(1),s_{F}(2),0,s_{F}(3)]^T$ & $\{2,4\}$ & $[0,s_{N}(1),0,s_{N}(2)]^T$ \\
		$[1,0,1]$ & $\{1,2,4\}$ & $[s_{F}(1),s_{F}(2),0,s_{F}(3)]^T$ & $\{1,2\}$ & $[s_{N}(1),s_{N}(2),0,0]^T$ \\
		$[1,1,0]$ & $\{1,2,3\}$ & $[s_{F}(1),s_{F}(2),s_{F}(3),0]^T$ & $\{2,3\}$ & $[0,s_{N}(1),s_{N}(2),0]^T$ \\
		$[1,1,1]$ & $\{1,2,3\}$ & $[s_{F}(1),s_{F}(2),s_{F}(3),0]^T$ & $\{1,2\}$ & $[s_{N}(1),s_{N}(2),0,0]^T$ \\ \hline \hline
	\end{tabular}
\end{table*}

The vector of the modulated symbols at $K_u$ active subcarriers for the user $u$ is given by 
\begin{align}
	{\bf s}_u = [s_{u}(1), ..., s_{u}(K_u)]^T,
\end{align}
where $s_{u}(\gamma) \in {\cal S}_u$, $\gamma \in \{1, 2,...,K_u\}$ and ${\cal S}_u$ is the complex signal constellation of size $M_u$. 

For high spectral efficiency and additional flexibility, we introduce the {\it Envelope Former} module to the transmitter. In this module, the indices of the active subcarriers are selected for each user according to the mixed data bits of FU and NU. 
Specifically, in each subblock, data bits for NU are added behind FU in the bit mixer to form one data stream. As we have discussed, the length of the data stream to form the envelope is
\begin{align}
	{m}_{index} = \left\lfloor {\log_2 \binom{L}{K_{F}}} \right\rfloor+\left\lfloor {\log_2 \binom{K_{F}}{K_{N}}} \right\rfloor,
\end{align}
bits. The subcarrier index selector utilizes the first $\left\lfloor {\log_2 \binom{L}{K_{F}}} \right\rfloor$ bits to decide the subcarrier activation pattern of the FU subblock and the remaining bits are for the NU subblock. Then the OFDM subblock is produced by the subcarrier mapper, which can be expressed as follows
\begin{align}
	{\bf x}_{u} = [x_{u}(1),x_{u}(2),...,x_{u}(L)]^T ,
\end{align}
where $x_{u}(\gamma) \in \{0,{\cal S}\}$ and $\gamma \in \{1,2,...,L\}$. The indices set of the active subcarriers is expressed as
\begin{align}
	{\bf I}_{u} = \{I_{u}(1),I_{u}(2),...,I_{u}(K_u)\},
\end{align}
where $I_{u}(1)<I_{u}(2)<...<I_{u}(K_u)$ and $I_u(k) \in \{1,2,...,L\}$.
An example of the mapping table is presented in Table \ref{table: mapping} for $L=4$, $K_{F}=3$, and $K_{N}=2$.

Afterwards, $G$ OFDM subblocks of user $u$ are concatenated together to form one OFDM block before the power allocation. To distinguish signals of two users, different transmit power levels $P_{N}$ and $P_{F}$ are allocated to NU and FU, respectively, and $P_{N}+P_{F}=P_{\rm max}$. Since the channel gain for the NU is higher than the FU due to the distance difference, a higher power level is allocated to FU for user fairness, i.e. $P_F > P_N$. We denote $P_u = a_uP_{max}$ and $\alpha_u = \sqrt{P_u}$. After the superposition coding, the tagged subblock can be expressed as
\begin{align}
	{\bf x} = \sum_{u} \alpha_u{\bf x}_u.
\end{align}

Later, the inverse fast Fourier transform (IFFT) is applied to obtain the time domain OFDM block. A cyclic prefix (CP) of length $Q$ is added at the beginning of the OFDM block and then the signal is transmitted to the users.

\vspace{-0.2 cm}
\subsection{Data Detection at Receivers with Decoding Scheme}
Here the data detection procedure for both users is presented. At the receiver, the fast Fourier transform is applied after removing the CP. The frequency domain signal vector received by the user $u$ is expressed as follows
\begin{align}
	{\bf y}_u = {\bf X}{\bf h}_u + {\bf w},
\end{align}
where ${\bf h}_u=[h_u(1),h_u(2),...,h_u(L)]^T$ and ${\bf w}=[w(1),w(2),...,w(L)]^T$ denote the channel vector and the additive white Gaussian noise vector, respectively. For $\gamma \in \{1,2,...,L\}$, $w(\gamma) \sim {\cal CN} \left(0,N_0 \right)$ and $h_u(\gamma) \sim \Omega_u{\cal CN} \left(0,1 \right)$, where $\Omega_u$ is the average channel gain. Usually, $\Omega_N > \Omega_F$ holds in NOMA systems.

The SIC is employed as in conventional NOMA systems. Without loss of generality, we assume that the SIC is always processed at the NU. Therefore, the NU has to detect and subtract the FU signal before decoding its message. The SIC detection of conventional OFDM-NOMA in a subcarrier-by-subcarrier manner is not feasible due to the envelope information conveyed by subblock realizations, and hence the NOMA-IE subblocks must be processed as a whole for detection. Therefore, we employ an envelope-detected SIC (E-SIC) in NOMA-IE. In the E-SIC procedure, the NU performs a joint search for all possible combinations of the activation patterns and the constellation symbols according to the mapping table. The detected FU subblock can be given by
\begin{align}
	{\hat {\bf x}}_{F,SIC} = \mathop{\arg\min}_{{\bf x}_F \in {\bf \chi}_F} \sum_{\gamma=1}^{L} \left| {y_N(\gamma) - \beta_e \alpha_F {\tilde h}_N(\gamma)} x_F(\gamma) \right|^2 ,
\end{align}
where $\beta_e >0$ is the envelope detection coefficient. In conventional SIC procedure, $\beta_e = 1$. $\chi_F$ stands for the possible realization set of the FU subblock ${\bf x}_F$. ${\tilde {\bf h}}_N = [{\tilde h}_N(1),{\tilde h}_N(2),...,{\tilde h}_N(L)]^T$ is the estimated channel vector at the NU. In this work, the perfect channel estimation is assumed. Note that the index bits of the FU signal can be decoded by the NU, the NU is able to use at most ${m}_{index}$ bits of index information without extra detection complexity. 

After the E-SIC procedure, the subblock is given by ${\bf y}_{N,SIC} = {\bf y}_N - \alpha_F {\hat {\bf X}}_{F,SIC} {\tilde {\bf h}}_N$. By utilizing the knowledge of active subcarrier indices of the FU signal, i.e., $\hat{\bf I}_{F,SIC}$, the NU is able to detect both the index bits and the symbol bits from these indices by the envelope-detected detection (E-detection). Similar to the E-SIC, the E-detection procedure detects the data bits on the envelope and modulated constellations simultaneously. The detected NU subblock is
\begin{align}
	{\hat {\bf x}}_{N} = \mathop{\arg\min}_{{\bf x}_N \in {\bf \chi}_N} \sum_{\gamma \in \hat{\bf I}_{F,SIC}} \left| {y_{N,SIC}(\gamma) - \alpha_N {\tilde h}_N(\gamma)} x_N(\gamma) \right|^2 ,
\end{align}
where $\chi_N$ represents the possible realization set of the NU subblock ${\bf x}_N$ conditioned on $\hat{\bf I}_{F,SIC}$.

Different from the NU, the FU decodes its message directly, treating the NU signal as interference. The FU subblock is detected by
\begin{align}
	{\hat {\bf x}}_{F} = \mathop{\arg\min}_{{\bf x}_F \in {\bf \chi}_F} \sum_{\gamma=1}^{L} \left| {y_{F}(\gamma) - \beta_e\alpha_F {\tilde h}_F(\gamma)} x_F(\gamma) \right|^2 ,
\end{align}
where ${\tilde {\bf h}}_F = [{\tilde h}_F(1),{\tilde h}_F(2),...,{\tilde h}_F(L)]^T$ is the estimated channel vector at the FU. It should be noted that an extra detection procedure is needed if the FU uses the index information of the NU signal.

\section{Performance Metrics and Prelimilary Analysis}
In this section, we provide the performance metrics to evaluate the proposed NOMA-IE scheme in this work, and then different OFDM-based schemes are compared. Afterwards, some fundamental properties of the NOMA-IE are discussed.

\subsection{Performance Evaluation Metrics}\label{sec: BER}
\subsubsection{Spectral Efficiency}
The spectral efficiency is defined as the ratio of the number of data bits in transmission to the total number of OFDM subcarriers in the considered OFDM block \cite{IMsurvey}, which is given by
\begin{align}
	{\rm SE} = \frac{N_t/L\times \sum_u m_u }{N_t+Q} \mbox{ bits/s/Hz},
\end{align}
where $m_{data} = \sum_u m_u$ is the overall transmitted data bits in an OFDM subblock and $u \in \{N,F\}$.

\subsubsection{Energy Efficiency}
In this work, we regard the ratio of transmitted data bits to the total average energy consumption as energy efficiency. Here the CP is ignored for simplicity. Two transmit-power management policies are considered as follows.

(i) {\it Maximum Transmit Power Policy:}
The maximum power of each subcarrier is limited by $P_{max}$. The energy efficiency is expressed as
\begin{align}
	{\rm EE} = \frac{\sum_u m_u }{P_{max}\sum_u a_uK_u } \mbox{ bits/J}.
\end{align}

(ii) {\it Power Reallocation Policy:}
As in \cite{IMpower}, the amplitude of each subcarrier is scaled by $\sqrt{\frac{L}{\sum_u a_uK_u}}$, and the energy efficiency is given by
\begin{align}
	{\rm EE} = \frac{\sum_u m_u }{LP_{max}} \mbox{ bits/J}.
\end{align}

\subsubsection{Bit Error Rate}
The BER is the ratio of the number of data bits in error to the total of data bits transmitted by the OFDM subblock. Since the envelope of the NOMA-IE is able to convey data bits, error bits in NOMA-IE result from two independent cases: 

(i) the indices of the active subcarriers are correct. In this case, the index bits are in error while the symbol bits have a probability to be correct;

(ii) the indices of the active subcarriers are correct, and only symbol bits are incorrect.  

We employ $m_{e1,g,u}$ and $m_{e2,g,u}$ to denote the number of error bits in subblock $g$ for these two cases, respectively. Note that the index bits can be ``borrowed" from or ``lent" to the other NOMA user as we have discussed, error bits $\Delta m_{e1,g,u}$ and transmitted error bits $\Delta m_u$ are considered. Then the BER for user $u$ can be expressed as
\begin{align}
	P_{b,u} = \frac{\sum_g \left( m_{e1,g,u}+m_{e2,g,u} + \Delta m_{e1,g,u}  \right) }{N_t/L \left( m_u + \Delta m_u \right)}.
\end{align}

The error performance of the NOMA-IE is investigated detailedly in the next section.

\begin{table*}[t!]
	\centering
	\caption{Performance Comparisons of Different OFDM-Based Schemes} \label{table: performance}	
	\begin{tabular}{
			|m{2 cm}<{\centering} |m{5.2cm}<{\centering}
			|m{5.2 cm}<{\centering}| }
		\hline
		Scheme & Spectral efficiency & Energy efficiency \\ \hline
		Conventional OFDM & $\frac{N_T \log_2M}{ N_T+Q  } $ & $\frac{\log_2M}{P_{max}}$ \\ \hline
		OFDM-NOMA \cite{OFDMNOMA} & $\frac{N_T\sum_u \log_2M_u}{ N_T+Q  } $ & $\frac{\sum_u \log_2M_u}{P_{max}}$ \\ \hline
		OFDM-IM \cite{IMOFDMBasar} & $\frac{N_T \left( \left\lfloor {\log_2 \binom{L}{K}} \right\rfloor + K\log_2M \right)}{ L\left(N_T+Q \right) } $ & $\frac{\left\lfloor {\log_2 \binom{L}{K}} \right\rfloor + K\log_2M}{KP_{max}}$ \\ \hline
		IM-NOMA \cite{IMNOMA} & $\frac{N_T \sum_u \left( \left\lfloor {\log_2 \binom{L}{K_u}} \right\rfloor + K_u\log_2M_u \right)}{ L\left(N_T+Q \right) } $ & $\frac{\sum_u \left( \left\lfloor {\log_2 \binom{L}{K_u}} \right\rfloor + K_u\log_2M_u \right)}{ P_{max} \sum_ua_uK_u} $  \\ \hline
		NOMA-IE (this paper) & $\frac{N_T \left( \left\lfloor {\log_2 \binom{L}{K_F}} \right\rfloor + K_F\log_2M_F \right)}{ L\left(N_T+Q \right) } + \frac{N_T \left( \left\lfloor {\log_2 \binom{K_F}{K_N}} \right\rfloor + K_N\log_2M_N \right)}{ L\left(N_T+Q \right) } $ & $   \frac{ \left\lfloor {\log_2 \binom{L}{K_F}} \right\rfloor + K_F\log_2M_F }{P_{max} \sum_ua_uK_u} + \frac{ \left\lfloor {\log_2 \binom{K_F}{K_N}} \right\rfloor + K_N\log_2M_N }{P_{max} \sum_ua_uK_u}  $ \\ \hline
	\end{tabular}
\end{table*}
\subsection{Prelimilary Analysis}
We compare the spectral efficiency and energy efficiency in the maximum transmit power policy of different OFDM-based schemes in Table \ref{table: performance}. In the power reallocation policy, the energy consumptions of all the considered schemes are equal, so we do not list the energy efficiency of this policy in the table. In those OMA schemes (including conventional OFDM and OFDM-IM), we use $K$ and $M$ to represent the number of active subcarriers in each subblock and the order of APM, respectively. Note that two users are grouped in NOMA schemes, and paired users are considered in OMA schemes for a fair comparison.  

It is obvious from the results that the NOMA improves spectral efficiency. By appropriately selecting $K_u$, the spectral efficiency performance of IM-NOMA and NOMA-IE are able to be better than the OFDM-NOMA. The NOMA-IE is also highly energy-efficient in the maximum transmit power policy because the OFDM-IM framework enables the OFDM waveform to transmit more data bits by the same number of active subcarriers than the conventional OFDM.

Moreover, an interesting interplay can be found between the OFDM-IM and the NOMA. On the one hand, the OFDM-IM brought flexibility to NOMA users. Utilizing the fact that the NU is able to use the FU index bits without extra detection complexity, the overall index bits $m_{index}$ can be reallocated with various proportions to users. This feature helps to meet different user requirements. On the other hand, the NOMA also helps the envelope of OFDM-IM to convey more information. Specifically, when considering a two-user NOMA group, the upper-bound bits of information conveyed by the envelope of NOMA-IE in a subblock are given by
\begin{align}
	m_{index} &= \log_2 \left(\sum_{K_F=0}^L\binom{L}{K_F} \sum_{K_N=0}^{K_F}\binom{K_F}{K_N} \right) \nonumber \\
	& = \log_2 \left(\sum_{K_F=0}^L\binom{L}{K_F} 2^{K_F} \right) = \log_2 \left( 3^L\right)  \mbox{\rm bits},
\end{align}
which is ternary. This conclusion is easily extended to the $N$-user NOMA group that the $(N+1)$-ary signal can be conveyed by the envelope of the NOMA-IE. In the OFDM-IM, the index information is binary.

\vspace{-0.4 cm}
\section{Error Performance Analysis}
In this section, we employ the BPSK modulation. To characterize the error performance, we first focus on the error probability that ${\bf x}_F$ is incorrectly detected as ${\hat {\bf x}}_F$ conditioned on known channel statement information (CSI) for the FU, which is expressed as
\begin{align}\label{eq: ep FU}
	{\rm Pr}\left({\bf x}_F \to {\hat {\bf x}}_F | {\tilde {\bf h}}_F \right) = {\rm Pr} \left(  \bigcap_{{\bf z}_F \in \chi_F \backslash {\hat {\bf x}}_F} \! \left\{  \left\| {\bf y}_F - \beta_e\alpha_F {\bf Z}_F{\tilde {\bf h}}_F  \right\|^2 > \left\| {\bf y}_F - \beta_e\alpha_F {\hat {\bf X}}_F {\tilde {\bf h}}_F \right\|^2 \right\} \right).
\end{align}

Similarly, for the NU, the conditional error probabilities of the SIC procedure and the NU signal detection are 
\begin{align}\label{eq: ep SIC}
	{\rm Pr}\left({\bf x}_F \to {\hat {\bf x}}_F | {\tilde {\bf h}}_N \right) = {\rm Pr} \left( \! \bigcap_{{\bf z}_F \in \chi_F \backslash {\hat {\bf x}}_N} \!\! \left\{  \left\| {\bf y}_N - \beta_e\alpha_F {\bf Z}_F{\tilde {\bf h}}_N  \right\|^2 > \left\| {\bf y}_N - \beta_e\alpha_F {\hat {\bf X}}_F {\tilde {\bf h}}_N \right\|^2 \right\} \right),
\end{align}
\begin{align}\label{eq: ep NU}
	{\rm Pr}\left({\bf x}_N \to {\hat {\bf x}}_N | {\tilde {\bf h}}_N \right) = {\rm Pr} \left( \! \bigcap_{{\bf z}_N \in \chi_N \backslash {\hat {\bf x}}_N} \! \left\{ \left\| {\bf y}_{N,SIC} - \alpha_N {\bf Z}_N{\tilde {\bf h}}_N  \right\|^2 > \left\| {\bf y}_{N,SIC} - \alpha_N {\hat {\bf X}}_N {\tilde {\bf h}}_N \right\|^2 \right\} \right).
\end{align}

It is difficult to calculate the exact probability when the number of possible realizations in $\chi_u$ is larger than two. As in most existing works, we utilize the pairwise error probability (PEP) as an approximation to simplify our derivations\cite{PEP}.
\begin{approximation}\label{approx: 1}
The PEP is the tight upper bound of the error probability in \eqref{eq: ep FU}, \eqref{eq: ep SIC}, and \eqref{eq: ep NU}. The conditional PEP expressions are given by
\begin{align}
	{\rm Pr}\left({\bf X}_F \to {\hat {\bf X}}_F | {\tilde {\bf h}}_F \right) \approx {\rm Pr} \left( \left\| {\bf y}_F - \beta_e\alpha_F {\bf X}_F{\tilde {\bf h}}_F  \right\|^2 > \left\| {\bf y}_F - \beta_e\alpha_F {\hat {\bf X}}_F {\tilde {\bf h}}_F \right\|^2 \right),
\end{align}
\begin{align}
	{\rm Pr}\left({\bf X}_F \to {\hat {\bf X}}_F | {\tilde {\bf h}}_N \right) \approx {\rm Pr} \left( \left\| {\bf y}_N - \beta_e\alpha_F {\bf X}_F{\tilde {\bf h}}_N  \right\|^2 > \left\| {\bf y}_N - \beta_e\alpha_F {\hat {\bf X}}_F {\tilde {\bf h}}_N \right\|^2 \right),
\end{align}
\begin{align}
	{\rm Pr}\left({\bf X}_N \to {\hat {\bf X}}_N | {\tilde {\bf h}}_N \right) \approx {\rm Pr} \left( \left\| {\bf y}_{N,SIC} - \alpha_N {\bf X}_N{\tilde {\bf h}}_N  \right\|^2 > \left\| {\bf y}_{N,SIC} - \alpha_N {\hat {\bf X}}_N {\tilde {\bf h}}_N \right\|^2 \right),
\end{align}
which converge at high SNR. Moreover, this approximation becomes exact when ${\bf x}_u$ is a symbol with the binary modulation.
\end{approximation}

Before our analysis of the NOMA-IE, we provide some simple conclusions in both the conventional OFDM and the OFDM-IM framework, which are helpful to derive expressions in the NOMA-IE. In conventional OFDM, symbols can be detected separately, so we only have to focus on one OFDM subcarrier. Since the NOMA-IE is a mixture of NOMA and OMA schemes, here we consider both the OMA and the NOMA cases. Noticed that the FU detects the superimposed signal directly, we denote the amplitude of the superimposed symbol as $|x(\gamma)| \triangleq \lambda_F \in {\bf \Lambda} = \{\lambda_-, \lambda_*, \lambda_+\}$ and the elements in ${\bf \Lambda}$ can be given by
\begin{subequations}
	\begin{align}
		\lambda_- = & \alpha_F-\alpha_N, \label{eq: lambdaa}\\
		\lambda_* &= \alpha_F, \label{eq: lambdab}\\
		\lambda_+ = & \alpha_F+\alpha_N. \label{eq: lambdac}
	\end{align}
\end{subequations}

After the SIC procedure, the signal for the NU is OMA, and the amplitude of the NU symbol is $\lambda_N = \alpha_N$. Then the conditional PEP is expressed as follows.
\begin{proposition}
In conventional OFDM, the conditional PEP is
\begin{align}
	P_{C,u} (\lambda_u,{\tilde {\bf h}}_u)  = Q\left( \sqrt{ \frac{2 \left\| \lambda_u {\tilde {\bf h}}_u \right\|^2}{N_0}} \right),
\end{align}
where $Q(t) = \frac{1}{\pi} \int_{0}^{\pi/2} \exp{\left(-t^2/(2\sin^2x)\right)}dx$ is the Q-function.
\end{proposition}

In the OFDM-IM framework, we consider a simple two-subcarrier subblock where only one subcarrier is active. According to {\bf Definition \ref{def: IC-NOMA}}, all the possible realizations of a superimposed subblock are $\chi = \{[\lambda_F,0]^T, [-\lambda_F,0]^T, [0,\lambda_F]^T, [0,-\lambda_F]^T\}$. We assume ${\bf x} = [\lambda_F,0]^T$ due to the symmetry of modulation constellations. As we have discussed in Section \ref{sec: BER}, two categories of error exist in the OFDM-IM. Therefore, there are two cases for the PEP expression.
\begin{proposition}
In the OFDM-IM framework, if the index information is erroneously detected, i.e. the case (i) error occurs, the conditional PEP for the FU and the NU are separately given by
\begin{align}\label{eq: IM IF}
P_{I,F} (\lambda_F,{\tilde {\bf h}}_F)  = Q\left( \frac{\left( \lambda_F-\frac{1}{2}\beta_e\lambda_* \right) \left\| {\tilde h}_F(1) \right\|^2 + \frac{1}{2} \beta_e\lambda_* \left\| {\tilde h}_F(2) \right\|^2}{\sqrt{\frac{N_0}{2} \left( \left\| {\tilde h}_F(1) \right\|^2 + \left\| {\tilde h}_F(2) \right\|^2 \right)} } \right),
\end{align}
\begin{align}
	P_{I,N} (\lambda_N,{\tilde {\bf h}}_N)  = Q\left( \sqrt{\frac{\left\| \lambda_N {\tilde h}_N(1) \right\|^2 + \left\| \lambda_N{\tilde h}_N(2) \right\|^2}{2N_0} }  \right).
\end{align}

If the subblock only has a incorrect symbol, this is the case (ii) error. The conditional PEP can be expresssed as
\begin{align}\label{eq: IM S}
	P_{S,u} (\lambda_u,{\tilde {\bf h}}_u)  = Q\left( \sqrt{ \frac{2 \left\| \lambda_u {\tilde h}_u(1) \right\|^2}{N_0}} \right).
\end{align}

For the SIC procedure, we have $P_{I,SIC} (\lambda_F,{\tilde {\bf h}}_N) = P_{I,F} (\lambda_F,{\tilde {\bf h}}_N)$ and $P_{S,SIC} (\lambda_F,{\tilde {\bf h}}_N) = P_{S,F} (\lambda_F,{\tilde {\bf h}}_N)$.
\end{proposition}

It can be observed from \eqref{eq: IM S} that under OFDM-IM, the PEP expression of case (ii) error is the same as that in the conventional OFDM system. In this case, the error performance is unrelated to the channel condition of the inactive subcarrier and the envelope detection coefficient. However, if the case (i) error occurs, the error performance in the NOMA setup shown in \eqref{eq: IM IF} is much different from the conventional counterpart. In the following lemma, we illustrate cases where the intra-cluster interference in NOMA causes the error floor.

\begin{lemma}\label{lemma: efloor}
The detection error floor exists when $\alpha_F \le \frac{2\alpha_N}{2-\beta_e}$.
\end{lemma}
\begin{IEEEproof}
If the error floor exists, the conditional PEP is not zero when the transmit SNR $P_{max}/N_0 \to \infty$. According to the property of the Q-function, when $x \to +\infty$ we have $Q(x) \to 0$. It can be observed from \eqref{eq: IM IF} that $\lambda_F-\frac{1}{2}\beta_e\lambda_* > 0$ guarantees the conditional PEP is 0 when SNR becomes large. Then the proof is completed.
\end{IEEEproof}

\begin{corollary}
If the envelope detection coefficient $\beta_e = 1$, the error floor always exists when $\alpha_F \le 2\alpha_N$. That is to say, the feasible range of power allocation coefficients for the paired NOMA-IE users is $P_F > 4P_N$.
\end{corollary}

\begin{corollary}
If the envelope detection coefficient $\beta_e = \frac{\alpha_F-\alpha_N}{\alpha_F}$, $\alpha_F> \frac{2\alpha_N}{2-\beta_e}$ always holds. In this case, the error floor is eliminated. We denote
\begin{align}
	\beta_e^* = \frac{\alpha_F-\alpha_N}{\alpha_F},
\end{align}
as the feasible envelope detection coefficient.
\end{corollary}

\begin{figure*}[t!] 
	\centering
	\subfigure[OFDM-NOMA]{\includegraphics[width=5.9in]{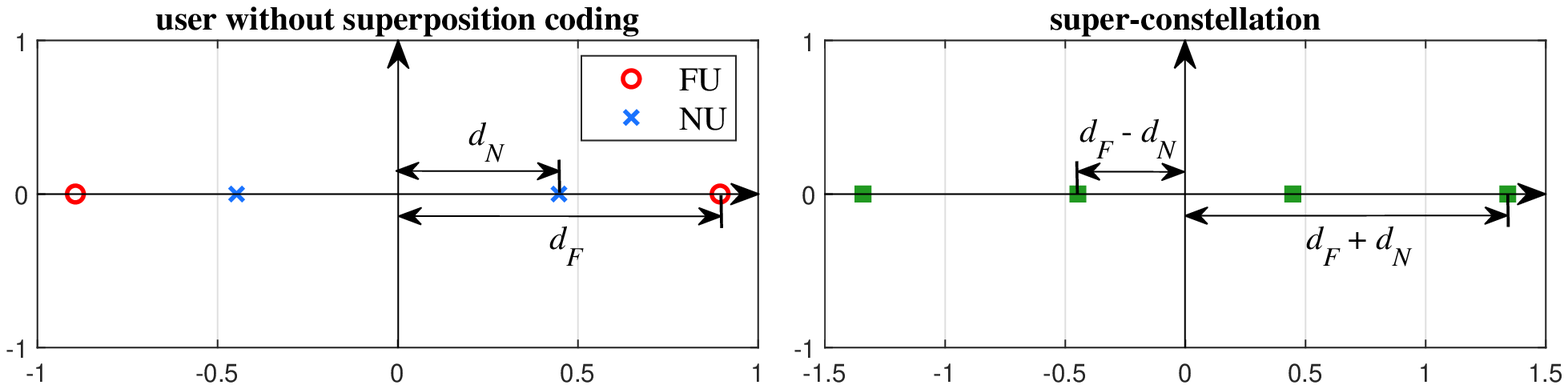}
		\label{con: fig_a}}
	\hfil
	\subfigure[IM-NOMA]{\includegraphics[width=5.9in]{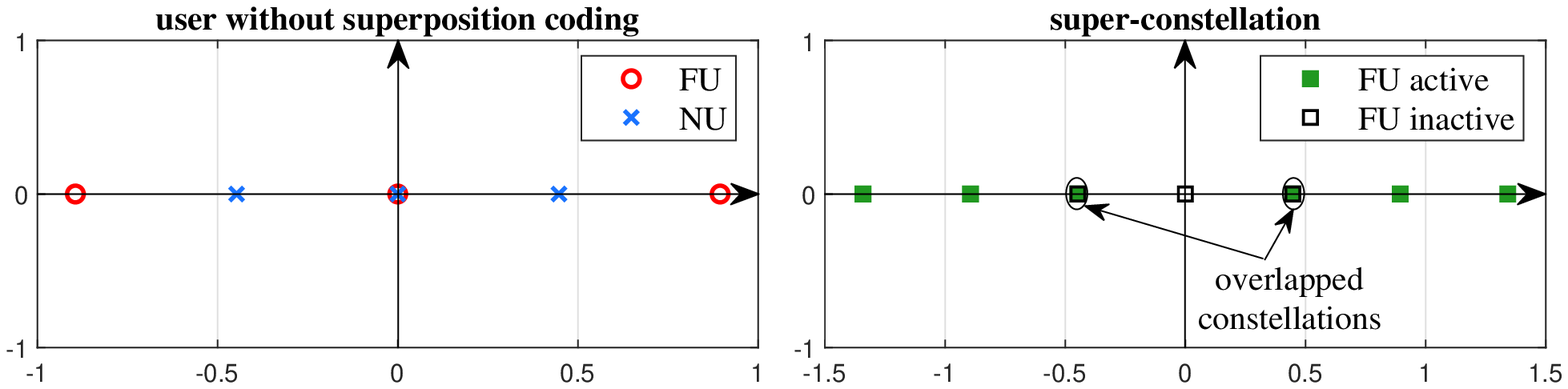}
		\label{con: fig_b}}
	\hfil
	\subfigure[NOMA-IE]{\includegraphics[width=5.9in]{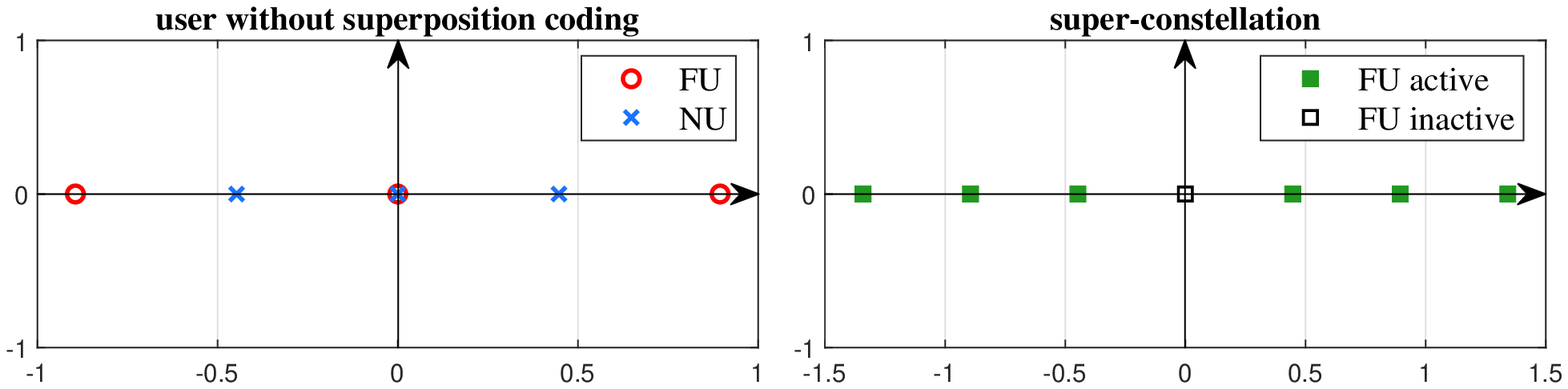}
		\label{con: fig_c}}
	\caption{The construction of the superimposed symbols with $\alpha_F = 2\alpha_N$.
	}
	\label{fig: constellation}
\end{figure*}

To understand the reason that results in the error floor more easily, we illustrate the construction of the superimposed symbols for the case $\alpha_F = 2\alpha_N$ in Fig. \ref{fig: constellation}, where the constellations with normalized energy are considered, i.e., $d_u = \frac{\alpha_u}{\sqrt{P_{max}}}$ for $u \in \{F,N\}$. It is observed that some superimposed constellation points cannot be distinguished in IM-NOMA when $\alpha_F \le 2\alpha_N$, which leads to the error floor no matter what value $\beta_e$ is. Different from the IM-NOMA, the design of NOMA-IE ({\bf Definition \ref{def: IC-NOMA}}) avoids the constellation overlap, and hence the appropriate $\beta_e$ helps to eliminate the error floor.

\begin{remark}
In NOMA-IE, the feasible range of power allocation coefficients for the paired NOMA-IE users is the same as that in conventional NOMA systems, i.e., $P_F > P_N$.
\end{remark}

%


By leveraging these conditional PEP expressions in the conventional OFDM and OFDM-IM frameworks, the theoretical BER for the NOMA-IE can be calculated. In the following subsections, we first provide the BER expressions in the two-subcarrier subblock, and then the results are extended to more complex multi-subcarrier subblocks. 

\subsection{BER in Two-subcarrier Subblock}
We consider $K_F = K_N = 1$ in this subsection. The FU subblock conveys 1 bit symbol bits and 1 bit index bits, while the NU subblock only conveys 1 bit symbol bits. If all the constellation points are equiprobable, the BER of the detected FU subblock at the FU can be expressed as
\begin{align}\label{eq: FBER2 1}
	P_{b,F} = \frac{1}{m_F \xi_F \xi_N} \sum_{{\bf x}_F} \sum_{{\bf x}_N} \sum_{{\hat {\bf x}}_F \ne {\bf x}_F} {\rm Pr}\left({\bf x}_F \to {\hat {\bf x}}_F \vert {\bf x}_N \right) e\left({\bf x}_F, {\hat {\bf x}}_F\right),
\end{align}
where $e\left({\bf x}, {\hat {\bf x}}\right)$ is the number of error bits when ${\bf x}$ is erroneously detected as ${\hat {\bf x}}$. $\xi_F$ and $\xi_N$ are the numbers of possible realizations for the FU subblock and the NU subblock, respectively.
Utilizing the symmetry of constellations, we assume that ${\bf x}_F =[1,0]^T $, and \eqref{eq: FBER2 1} can be rewritten as
\begin{align}\label{eq: FBER2 2}
	P_{b,F} &= \frac{1}{m_F \xi_N} \sum_{{\bf x}_N} \sum_{{\hat {\bf x}}_F \ne {\bf x}_F} {\rm Pr}\left({\bf x}_F \to {\hat {\bf x}}_F \vert {\bf x}_N \right) e\left({\bf x}_F, {\hat {\bf x}}_F\right) \nonumber \\
	&= \frac{1}{2m_F} \sum_{\lambda_F \in \{\lambda_-,\lambda_+\} } \left( P_{S,F}(\lambda_F) +3P_{I,F}(\lambda_F) \right),
\end{align}
where $P_{S,F}(\lambda)$ and $P_{I,F}(\lambda)$ are the unconditional PEP. Adopting $Q(x) \cong \frac{1}{12}e^{-x^2/2} + \frac{1}{4}e^{-2x^2/3}$ as the approximation of the Q-function \cite{Qfappr}, $P_{S,F}(\lambda)$ is given by
\begin{align}
	P_{S,F}(\lambda) &= \int_{{\tilde {\bf h}}_F}  Q\left( \lambda \sqrt{ \frac{2\left\| {\tilde {\bf h}}_F \right\|^2}{N_0}} \right) f\left( {\tilde {\bf h}}_F \right) d{\tilde {\bf h}}_F \nonumber \\
	& \cong \mathbb{E}_{\left\| {\tilde {\bf h}}_F \right\|^2} \left(\frac{1}{12}e^ {-\frac{\lambda^2 \left\| {\tilde {\bf h}}_F \right\|^2}{N_0}}+ \frac{1}{4} e^{-\frac{4\lambda^2 \left\| {\tilde {\bf h}}_F \right\|^2}{3N_0}} \right) \nonumber \\
	&=  \frac{1/12}{\Omega_F \lambda^2/ N_0 +1} + \frac{1/4}{4\Omega_F \lambda^2/ 3N_0 +1}.
\end{align}

The unconditional PEP $P_{I,F}(\lambda)$ is expressed as
\begin{align}\label{eq: IM I}
	P_{I,F}(\lambda) = \int_{0}^{\infty} \int_{0}^{\infty} Q\left( \Omega_F \frac{\left( \lambda-\frac{1}{2}\beta_e\lambda_* \right) h_1 + \frac{1}{2} \beta_e\lambda_* h_2}{\sqrt{\frac{N_0}{2} \left( h_1 + h_2 \right)} } \right)f(h_1)f(h_2) dh_1 dh_2 ,
\end{align}
where $f(x) = \exp(-x)$ due to the Rayleigh fading. 
When $a_F \to 1$, the interference from the NU is negligible, and hence the $P_{I,F}(\lambda)$ can be further simplified as
\begin{align}
	P_{I,F}(\lambda_F) &\cong P_{I,F}(\lambda_*) \nonumber \\ &= \frac{1/12}{\det \left( q_1\Omega_F {\bf A} +{\bf I}_2 \right)} + \frac{1/4}{\det \left(q_2\Omega_F {\bf A} +{\bf I}_2 \right)},
\end{align}
where ${\bf A}_F = ({\bf X}_F - {\hat{\bf X}}_F)^H({\bf X}_F - {\hat{\bf X}}_F) $, $q_1 = \frac{1}{4N_0}$, and $q_2 = \frac{1}{3N_0}$.

For the NU, the E-SIC process occurs first. The probability of the imperfect SIC conditioned on the known ${\bf x}_N$ can be expressed as
\begin{align}\label{eq: SIC2} 
	P_{e,SIC|{\bf x}_N} &= \sum_{{\hat {\bf x}}_{F,SIC} \ne {\bf x}_{F,SIC}} {\rm Pr}\left({\bf x}_{F,SIC} \to {\hat {\bf x}}_{F,SIC} \vert {\bf x}_N \right).
\end{align}
 
The calculation of \eqref{eq: SIC2} is the same as $P_{b,F}$. When the FU subblock is detected incorrectly, cases for the re-constructed signal ${\bf y}_{N,SIC}$ are complex.
Part of the NOMA interference improves the amplitudes of symbols while the other part of the interference degrades the error performance. 
 
For simplicity, we make an approximation to calculate the BER of the NU under imperfect SIC.
\begin{figure*}[t!] 
	\centering
	\subfigure[]{\includegraphics[width=2.9in]{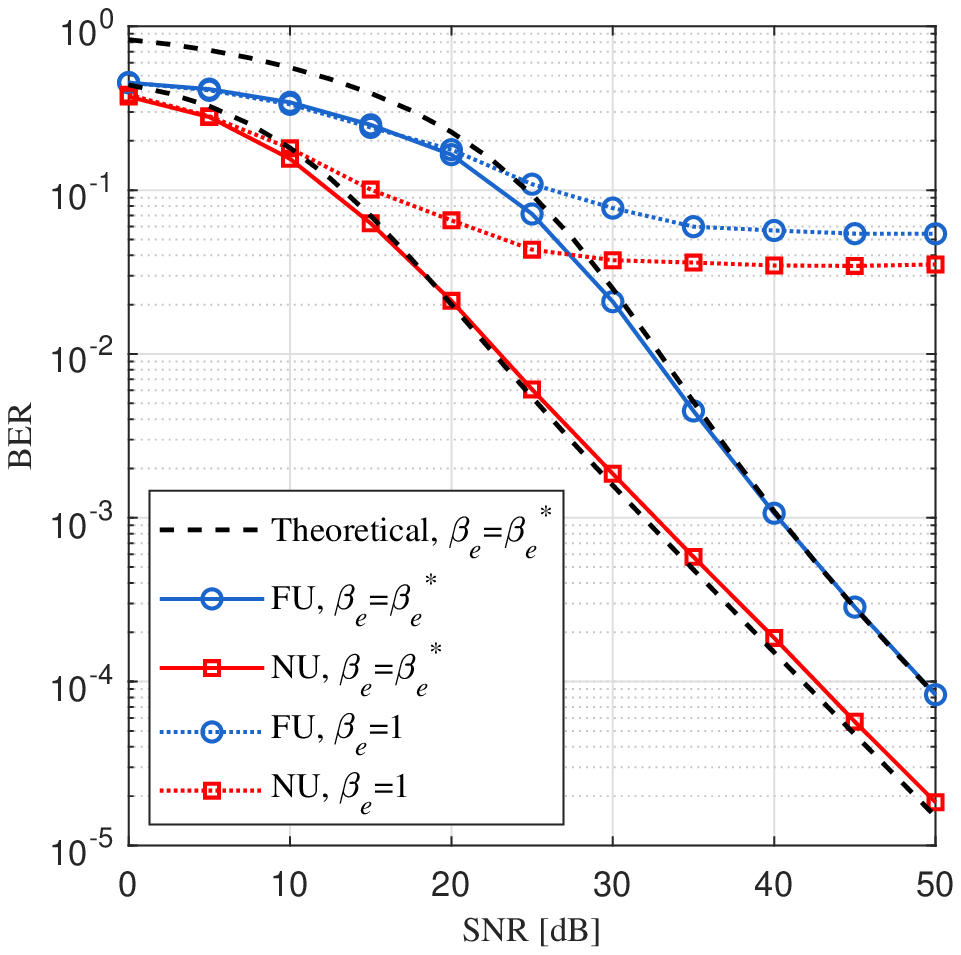}
		\label{d: fig_a}}
	\hfil
	\subfigure[]{\includegraphics[width=2.9in]{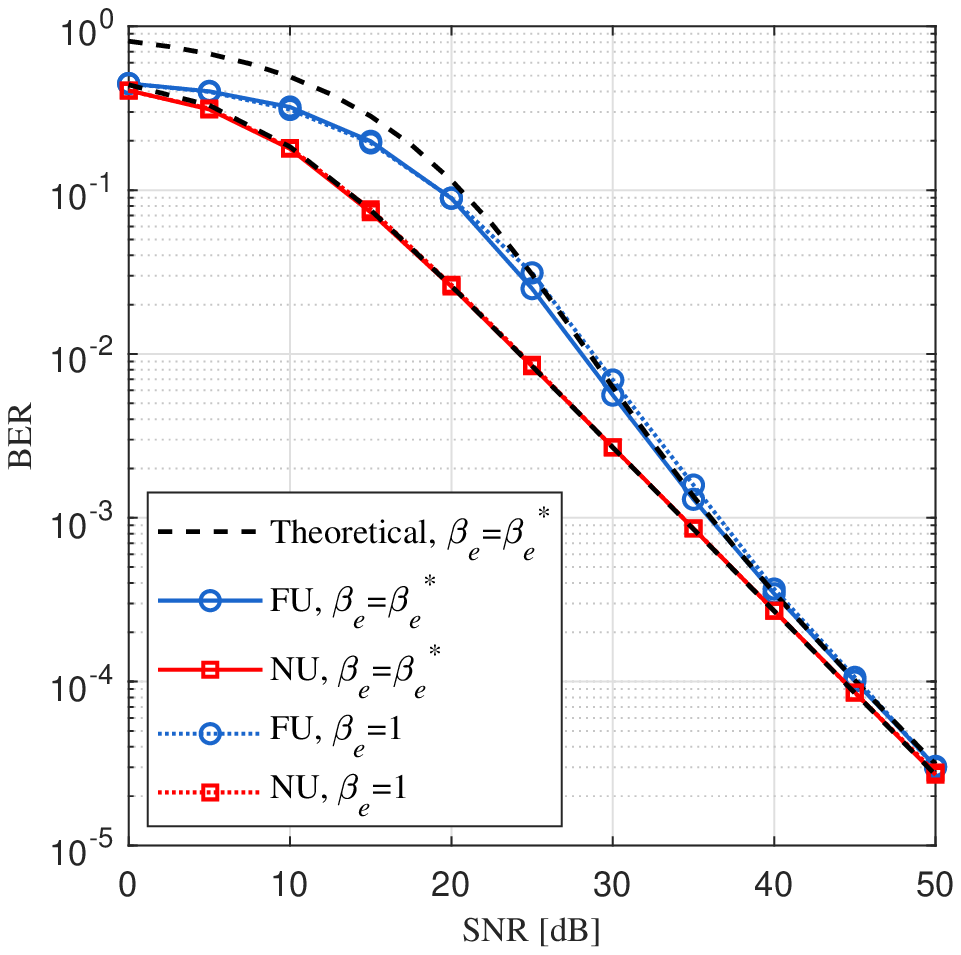}
		\label{d: fig_b}}
	\caption{Error performance of NOMA-IE with $L=2$, $K_F=K_N=1$ under different power allocation coefficients: (a) $a_F = 0.75$, $a_N = 0.25$; (b) $a_F = 0.9$, $a_N = 0.1$.
	}
	\label{fig: valid2}
\end{figure*}

\begin{approximation}\label{approx: 2}
Considering BPSK, the BER of the NU under imperfect SIC is 0.5. This approximation is based on the assumption that all the constellation points of the FU and the NU are equiprobable.
\end{approximation}
This approximation is accurate when $a_F \to 1$.
Then the BER of the NU subblock can be approximated by
\begin{align}\label{eq: NBER2 1}
	P_{b,N} &\approx \frac{1}{\xi_N}\sum_{{\bf x}_N} \left\{ 0.5P_{e,SIC|{\bf x}_N} +  \frac{1-P_{e,SIC|{\bf x}_N}}{m_N} \sum_{{\hat {\bf x}}_N \ne {\bf x}_N} {\rm Pr}\left({\bf x}_N \to {\hat {\bf x}}_N \right) e\left({\bf x}_N, {\hat {\bf x}}_N\right) \right\} \nonumber \\
	&= 0.5P_{e,SIC} + \frac{1}{2}\sum_{{\bf x}_N} \left\{  \left(1-P_{e,SIC|{\bf x}_N}\right) P_{C,N}(\lambda_N) \right\},
\end{align}
where
\begin{align}
	P_{C,N}(\lambda) &\cong \frac{1/12}{\Omega_N \lambda^2/ N_0 +1} + \frac{1/4}{4\Omega_N \lambda^2/ 3N_0 +1}.
\end{align}

From \eqref{eq: NBER2 1} we observe that the probability of the imperfect SIC $P_{e,SIC|{\bf x}_N}$ is a significant component of $P_{b,N}$. Therefore, a large $P_{e,SIC|{\bf x}_N}$ leads to poor error performance for the NU.
\begin{remark}
A low BER of the NU in NOMA-IE heavily relies on the perfect SIC. This conclusion can be extended to the general NOMA transmissions with the SIC procedure. 
\end{remark}

\begin{figure} [t!]
	\centering
	\includegraphics[width = 3.4in] {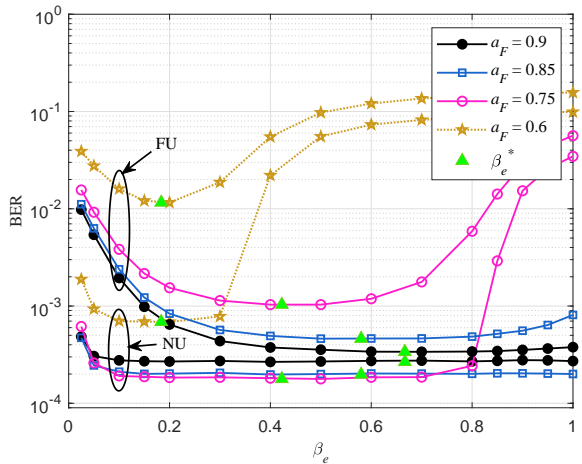}
	\caption{BER versus envelope detection coefficient $\beta_e$ with different power allocation coefficients.}
	\label{fig: betathres}
\end{figure}

In Fig. \ref{fig: valid2}, we present simulation results to validate the theoretical expressions of the BER in \eqref{eq: FBER2 2} and \eqref{eq: NBER2 1}, where $\Omega_F= -6$ dB and $\Omega_N= 0$ dB. For the FU, theoretical curves converge to simulation ones with the increase of SNR. For the NU, there is a small gap between the theoretical and the simulation results because {\bf Approximation 2} is made for the case of imperfect SIC. As we have discussed in {\bf Lemma \ref{lemma: efloor}}, the error floor exists in conventional detection procedure ($\beta_e = 1$) when $a_F = 0.75$. By adjusting the envelope detection coefficient to $\beta_e^* = \frac{\alpha_F - \alpha_N}{\alpha_F}$ ({\bf Corollary 2}) the error floor can be avoided.

To further show the impact of the value of $\beta_e$ on the error performance, we plot the BER of two users versus $\beta_e$ with different power allocation coefficients in Fig. \ref{fig: betathres}. In general, with the increase of $\beta_e$, the BER performance is improved to the best first and then decreases. When $a_F \le 0.8$, the BER significantly degrades at $\beta_e = 1$. It can be observed that the feasible envelope detection coefficient $\beta_e^*$ always helps to keep the BER at a low level in the NOMA-IE scheme.

\begin{remark}
In two-user NOMA-IE and BPSK considered, the high error performance is obtained by employing the feasible envelope detection coefficient $\beta_e^*$.
\end{remark}

\subsection{BER in Four-subcarrier Subblock}
We consider $K_F = 3$ and $K_N = 2$ shown in Table \ref{table: mapping} as an example. Similarly, we first focus on the FU. Based on {\bf Approximation \ref{approx: 1}}, the unconditional PEP of the FU subblock
\begin{align}
	{\rm Pr}\left({\bf x}_F \to {\hat {\bf x}}_F \right) = \int_{{\tilde {\bf h}}_F} {\rm Pr} \left( \left\| {\bf y}_F - \beta_e\alpha_F {\bf x}_F{\tilde {\bf h}}_F  \right\|^2 > \left\| {\bf y}_F - \beta_e\alpha_F {\hat {\bf x}}_F {\tilde {\bf h}}_F \right\|^2 \right) d {\tilde {\bf h}}_F,
\end{align}
has at least quadruple integral. The expression seems tedious and intractable.

Fortunately, the four-subcarrier superimposed subblock can be regarded as the combination of multiple two-subcarrier subblocks. We call them \emph{two-subcarrier elements (TSEs)}. The number of active subcarriers in TSEs is $K_E = 1$ and we denote the $k$th TSE as $T_{F,k} = \{ c_F(k),0\}$ as the FU message is always detected first. In particular, the superimposed subblock consists of $K_F$ TSEs, and all these elements share the same inactive subcarrier. The amplitude vector of symbols on active subcarriers is denoted by ${\bf c}_F= [c_F(1),...,c_F(K_F)]^T \in \mathbb{C}^{K_F \times 1}$, where $c_F(k) \in \{\lambda_-, \lambda_*, \lambda_+\}$. According to our definition, $c_F(k)$ is equal to the amplitude of $x_F(I_F(k))$, i.e., $c_F(k) = |x_F(I_F(k))|$. 

If the FU subblock can be detected correctly, no error occurs in any TSEs. Therefore, the subblock correct probability can be expressed as
\begin{align}
	P_{BL,c} = \int_{0}^{\infty} \exp(-x) \prod_{k=1}^{K_F} \left( 1-2P_{I,F} (c_F(k),x) - P_{S,F} (c_F(k))\right) d x,
\end{align}
where
\begin{align}
	P_{I,F}(\lambda,x) = \int_{0}^{\infty} Q\left( \Omega_F \frac{\left( \lambda-\frac{1}{2}\beta_e\lambda_* \right) h_1 + \frac{1}{2} \beta_e\lambda_* x}{\sqrt{\frac{N_0}{2} \left( h_1 + x \right)} } \right)f(h_1) dh_1.
\end{align}

On the other hand, at least one erroneous TSE leads to an incorrect FU subblock. We discuss two error cases for the FU subblock as follows.

\subsubsection{Case (i) Error}
In this case, the inactive subcarrier is erroneously detected as an active one. Therefore, the index error occurs in no less than one TSE. Since the probability that more than one TSE has the index error is ignorable at a high SNR, we employ the following approximation.
\begin{approximation}\label{approx: 3}
The number of incorrect inactive indices is equal to the number of TSEs with the index error. This approximation converges the exact value when SNR$\to \infty$ in no error floor cases.
\end{approximation}

For clarity, we first investigate the index error probability. We define $\hat{{\bf I}}_F(k)$ as the indices set of active subcarriers when $k$th TSE has the index error. The probability of erroneously detecting ${\bf I}_F$ as $\hat{{\bf I}}_F(k)$ can be calculated by
\begin{align}
	{\rm Pr}\left({\bf I}_F \to \hat{{\bf I}}_F(k)\right) = \int_{0}^{\infty}  2P_{I,F} (c_F(k),x) \exp(-x) \prod_{j=1,j\ne k}^{K_F} \left( 1-2P_{I,F} (c_F(j),x)\right) d x.
\end{align}

According to the law of total probability, the probability that ${\bf s}_F$ is erroneously detected by $\hat{{\bf s}}_F$ can be expressed as
\begin{align}
	{\rm Pr}\left({\bf s}_F \to \hat{{\bf s}}_F| \hat{\bf I}_F \ne {\bf I}_F\right) = \sum_{k = 1}^{K_F} {\rm Pr}\left({\bf s}_F \to \hat{{\bf s}}_F | {\bf I}_F \to \hat{{\bf I}}_F(k) \right){\rm Pr}\left({\bf I}_F \to \hat{{\bf I}}_F(k)\right),
\end{align}
where ${\rm Pr}\left({\bf s}_F \to \hat{{\bf s}}_F | {\bf I}_F \to \hat{{\bf I}}_F(k) \right)$ represents the conditional PEP of detecting ${\bf s}_F$ as $\hat{{\bf s}}_F$ conditioned on ${\bf I}_F \to \hat{{\bf I}}_F(k)$. In the OFDM-IM framework, erroneous indices result in symbol unsynchronization. For example, if the FU subblock ${\bf x}_F = [s_F(1),s_F(2),s_F(3),0]^T$ is incorrectly detected as $\hat{{\bf x}}_F = [\hat{s}_F(1),0,\hat{s}_F(2),\hat{s}_F(3)]^T$, only $\hat{s}_F(1)$ is detected from $s_F(1)$. $\hat{s}_F(2)$ is detected from $s_F(2)$ and $\hat{s}_F(3)$ is from the noise. We denote ${\bf \Psi}_{{\bf I}_F \to \hat{{\bf I}}_F(k)}$ and $\bar{{\bf \Psi}}_{{\bf I}_F \to \hat{{\bf I}}_F(k)}$ as the synchronized symbol set and the unsynchronized symbol set conditioned on ${\bf I}_F \to \hat{{\bf I}}_F$, respectively, and we have ${\bf \Psi}_{{\bf I}_F \to \hat{{\bf I}}_F(k)} \bigcup \bar{{\bf \Psi}}_{{\bf I}_F \to \hat{{\bf I}}_F(k)} = {\bf I}_F$. The unsynchronized symbols can be detected as random constellation points of the employed modulation scheme. For those synchronized symbols, their error performance is related to channel conditions. We use ${\bf \Psi}_c$ and ${\bf \Psi}_e$ to denote the correct synchronized symbol set and the erroneous synchronized symbol set, respectively. Then the conditional PEP is given by
\begin{align}
	{\rm Pr}\left({\bf s}_F \to \hat{{\bf s}}_F | {\bf I}_F \to \hat{{\bf I}}_F(k) \right) = 0.5^{|\bar{{\bf \Psi}}_{{\bf I}_F \to \hat{{\bf I}}_F(k)}|} \prod_{I_F(i) \in {\bf \Psi}_c} P_{S,F} (c_F(i))\prod_{I_F(j) \in {\bf \Psi}_e} P_{S,F} (1-c_F(j)).
\end{align}

When the indices of the active subcarriers are erroneous, some symbols are detected from the random noise or unsynchronized symbols. Therefore, the symbol error probability is much higher than the case no index error occurs. Therefore, how to reduce the index error is an important issue for error performance improvement in the NOMA-IE.

\begin{remark}
In the OFDM-IM framework, the index error is strongly correlated with the symbol error, and the high probability of the index error leads to significant performance degradation in BER. 
\end{remark}

Afterwards, the number of error bits for the FU subblock in the index error case can be calculated as
\begin{align}
	m_{e1,F} = & \sum_{\hat{{\bf s}}_F \ne {\bf s}_F} {\rm Pr}\left({\bf s}_F \to \hat{{\bf s}}_F | \hat{\bf I}_F \ne {\bf I}_F \right)e\left({\bf s}_F \to \hat{{\bf s}}_F\right) \nonumber \\ &+ \sum_{\hat{{\bf I}}_F \ne {\bf I}_F} {\rm Pr}\left({\bf I}_F \to \hat{{\bf I}}_F(k)\right)e\left({\bf I}_F \to \hat{{\bf I}}_F(k)\right),
\end{align}
where $e\left({\bf s}_F \to \hat{{\bf s}}_F\right)$ is the number of error bits when ${\bf s}_F$ is erroneously detected as $\hat{\bf s}_F$. $e\left({\bf I}_F \to \hat{{\bf I}}_F(k)\right)$ is the number of error bits when ${\bf I}_F$ is erroneously detected as $\hat{\bf I}_F(k)$.

\subsubsection{Case (ii) Error}
In this case, the index information is correct, the probability of which can be calculated as
\begin{align}
	P_{I,c} = 1 - \int_0^\infty \exp(-x) \prod_{k=1}^{K_F} \left( 1-2P_{I,F} (c_F(k),x)\right) d x .
\end{align}

Since the indices of the active subcarrier are correct, all symbols in the FU subblock are synchronized. The PEP conditioned on the correct index information can be expressed as
\begin{align}
	{\rm Pr}\left({\bf s}_F \to \hat{{\bf s}}_F | \hat{\bf I}_F = {\bf I}_F \right) = \prod_{I_F(i) \in {\bf \Psi}_c} P_{S,F} (c_F(i))\prod_{I_F(j) \in {\bf \Psi}_e} P_{S,F} (1-c_F(j)).
\end{align}

Then the number of error bits for the FU subblock in the no index error case is given by
\begin{align}
	m_{e2,F} = P_{I,c}\sum_{\hat{{\bf s}}_F \ne {\bf s}_F} {\rm Pr}\left({\bf s}_F \to \hat{{\bf s}}_F | \hat{\bf I}_F = {\bf I}_F \right)e\left({\bf s}_F \to \hat{{\bf s}}_F\right).
\end{align}

As mentioned, no extra detection complexity is required if $\Delta m_N$ out of $\left\lfloor {\log_2 \binom{L}{K_{F}}} \right\rfloor$ bits index information of the FU subblock is used by the NU. When the NU utilizes the envelope of the FU subblock, the overall BER of the FU is
\begin{align}
	P_{b,F} = \frac{m_{e1,F}+m_{e2,F}+\Delta m_{e1,F}}{m_F-\Delta m_N},
\end{align}
where $\Delta m_{e1,F} = -p_1\sum_{\hat{{\bf I}}_F \ne {\bf I}_F} {\rm Pr}\left({\bf I}_F \to \hat{{\bf I}}_F(k)\right)e\left({\bf I}_F \to \hat{{\bf I}}_F(k)\right)$ and $p_1 = \Delta m_N / \left\lfloor {\log_2 \binom{L}{K_{F}}} \right\rfloor$.

At the NU, the E-SIC procedure is operated first. The probability of the imperfect SIC is given by
\begin{align}
	P_{SIC,e} = \int_{0}^{\infty} \exp(-x) \prod_{k=1}^{K_F} \left(2P_{I,SIC} (c_F(k),x) + P_{S,SIC} (c_F(k))\right) d x,
\end{align}
where
\begin{align}
	P_{I,SIC}(\lambda,x) = \int_{0}^{\infty} Q\left( \Omega_N \frac{\left( \lambda-\frac{1}{2}\beta_e\lambda_* \right) h_1 + \frac{1}{2} \beta_e\lambda_* x}{\sqrt{\frac{N_0}{2} \left( h_1 + x \right)} } \right)f(h_1) dh_1.
\end{align}

Conditioned on the perfect SIC, the calculation of the number of error bits for the NU subblock is similar to that for the FU subblock. The probability of the index error is $P_{I,N,e} = 2P_{I,N}(\alpha_N)$. The number of error bits for the case (i) error of the NU subblock can be easily obtained as
\begin{align}
	m_{e1,N} = &\sum_{\hat{{\bf s}}_N \ne {\bf s}_N} 0.5^{K_N}e\left({\bf s}_N \to \hat{{\bf s}}_N\right) +  P_{I,N,e}.
\end{align}

Considering the case (ii) erorr, the number of error bits is
\begin{align}
	m_{e2,N} = (1-P_{I,N,e})\sum_{i = 1}^{K_N} i \left(P_{S,N} (\alpha_N) \right)^{K_N-i} \left(1-P_{S,N} (\alpha_N) \right)^{i}.
\end{align}

If the SIC is imperfect, we adopt the approximation of the BER as in the two-subcarrier subblock. Similarly, the overall BER of the NU is approximated as
\begin{align}
	P_{b,N} \approx \frac{0.5m_NP_{e,SIC} + (1-P_{e,SIC})\left(m_{e1,N} + m_{e2,N} \right) + \Delta m_{e1,N}}{m_N+\Delta m_N},
\end{align}
where $\Delta m_{e1,N}$ is the number of error bits from the FU envelope under imperfect SIC. The calculation of $\Delta m_{e1,N}$ is similar to $\Delta m_{e1,F}$ and hence we skip it here. 

\vspace{-0.3 cm}
\section{Numerical Results}
In this section, numerical results are presented to validate the analysis of the four-subcarrier subblock and some interesting insights are provided. The BER performance of these schemes was evaluated through Monte Carlo simulations. The average channel gains for the FU and the NU are $\Omega_F = -6$ dB and $\Omega_N = 0$ dB, respectively. The feasible envelope detection coefficient $\beta_e^*$ is employed. We use FU$(m_F+\Delta m_F,L,K_F)$ and NU$(m_N+\Delta m_N,K_F,K_N)$ to denote the envelope-forming scheme. Unless otherwise stated, we set FU$(4,4,3)$ and NU$(4,3,2)$ for the equal spectral efficiency of the FU and the NU.
\begin{table}
	\centering
	\caption{The IM-NOMA Mapping Table for $L=4$ and $K_{u}=2$} \label{table: mappingIM}	
	\begin{tabular}{
			m{1.9cm}<{\centering} |m{1.6cm}<{\centering}
			|m{3cm}<{\centering} }
		\hline \hline
		Index bits & ${\bf I}_{u}$ & subblock ${\bf x}_u$ \\ \hline
		$[0,0]$ & $\{1,2\}$ & $[s_{u}(1),s_{u}(2),0,0]^T$ \\ 
		$[0,1]$ & $\{2,3\}$ & $[0,s_{u}(1),s_{u}(2),0]^T$  \\
		$[1,0]$ & $\{3,4\}$ & $[0,0,s_{u}(1),s_{u}(2)]^T$  \\
		$[1,1]$ & $\{1,4\}$ & $[s_{u}(1),0,0,s_{u}(2)]^T$  \\ \hline \hline
	\end{tabular}
\end{table}

\subsection{Benchmark Schemes}
To verify the effectiveness of the proposed NOMA-IE scheme, comparisons are made with conventional OFDM, OFDM-NOMA, and IM-NOMA in the existing work. The detailed settings are as follows.
\begin{itemize}
	\item {\bf OFDM-NOMA}: As in the considered NOMA-IE, BPSK modulation is employed for the NU and the FU. In particular, OFDM-NOMA can be regarded as a special case of the NOMA IE with $L = K_F = K_N$.
	\item {\bf IM-NOMA}: In this case, the activation patterns of the FU and the NU subblock are independent. We consider $L=4$, $K_F = K_N = 2$, and the BPSK modulation. The mapping table for the user $u \in \{F,N\}$ is shown in Table \ref{table: mappingIM}.
	\item {\bf OFDM}: In this case, the available bandwidth is equally allocated to two users. For a fair comparison with NOMA schemes with the BPSK modulation, 4 amplitude-shift keying (4ASK) modulation is employed, which guarantees the OMA scheme and NOMA schemes are of the same spectral efficiency.
\end{itemize}

\begin{figure*}[t!] 
	\centering
	\subfigure[]{\includegraphics[width=2.9in]{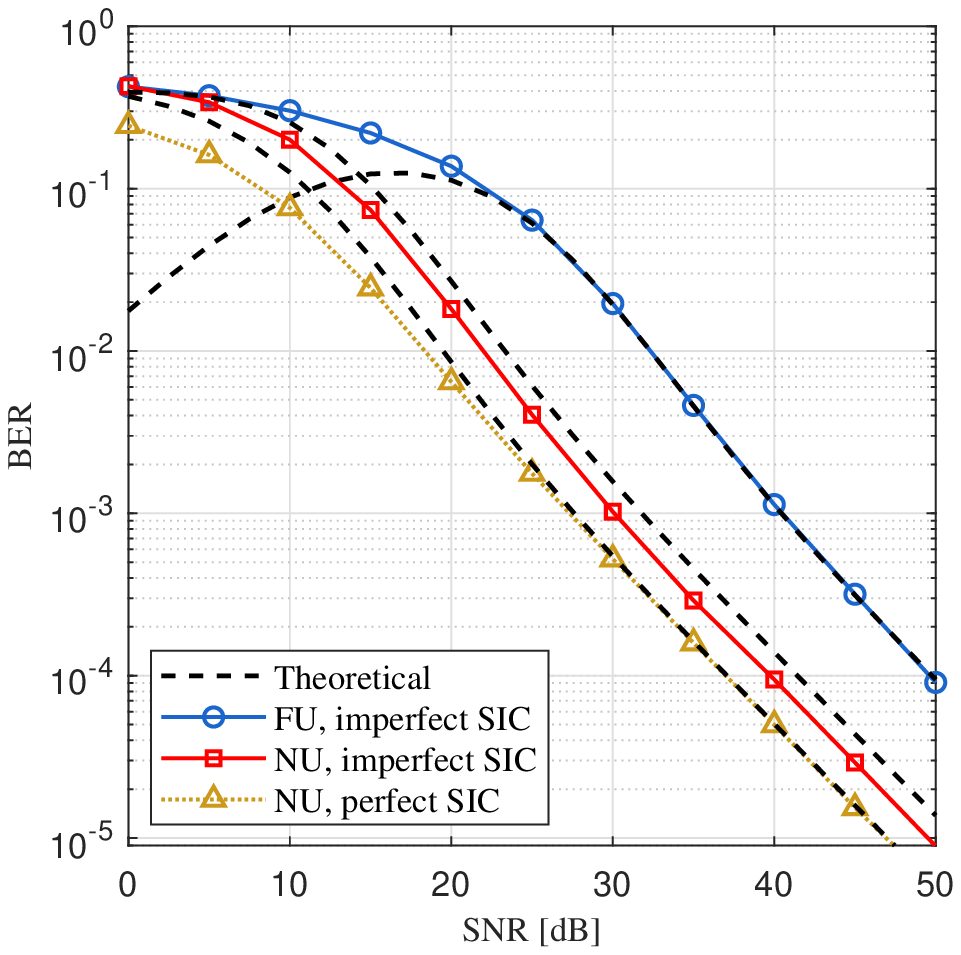}
		\label{f: fig_a}}
	\hfil
	\subfigure[]{\includegraphics[width=2.9in]{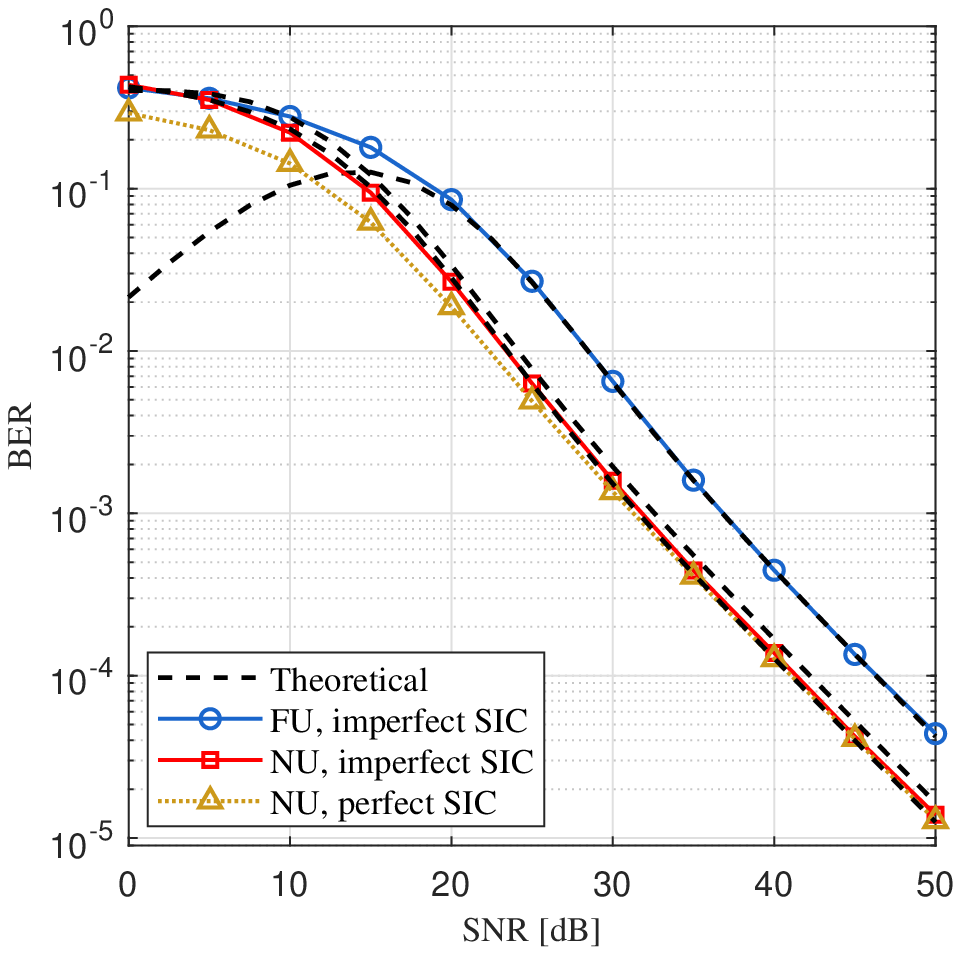}
		\label{f: fig_b}}
	\caption{Error performance of NOMA-IE with $L=4$, $K_F=3$, and $K_N=2$ under different power allocation coefficients: (a) $a_F = 0.75$, $a_N = 0.25$; (b) $a_F = 0.9$, $a_N = 0.1$.
	}
	\label{fig: valid4}
	\vspace{-0.6 cm}
\end{figure*}

\subsection{Validation of Theoretical Results}
\vspace{-0.2 cm}
Under different power allocation coefficients, the theoretical BER of the FU and the NU and the simulation results are illustrated in Fig. \ref{fig: valid4}, which validate the feasibility to express the theoretical BER of the multi-subcarrier subblock by the combination of TSEs. In the high SNR regime, the simulation curves of the FU fit the analytical results quite well. The observation verifies the asymptotic feature of {\bf Approximation \ref{approx: 1}} and {\bf Approximation \ref{approx: 3}}. For the NU, we consider the perfect SIC case for comparison. It can be observed that the theoretical result of the perfect SIC case is accurate, while a gap between the simulation and the analytical curves exists even when the SNR is high in the imperfect SIC case. This gap is from {\bf Approximation \ref{approx: 2}} and can be bridged with the increase of the power allocation coefficient $\alpha_F$ as we have discussed. Moreover, comparing Fig. \ref{f: fig_a} with Fig. \ref{f: fig_b}, the BER degradation from the imperfect SIC becomes serious with a small $\alpha_F$. This is because the probability of imperfect SIC grows when power allocated to the FU decreases, while the error performance is significantly affected by the imperfect SIC.

\vspace{-0.4 cm}
\subsection{Performance of Different Envelope-Forming Schemes}
\begin{figure} [t!]
	\centering
	\includegraphics[width = 3.4in] {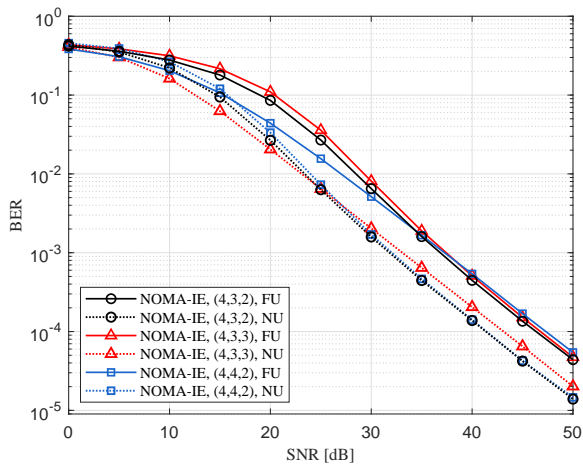}
	\caption{Error performance versus SNR in different envelope-forming schemes.}
	\label{fig: env}
	\vspace{-0.3 cm}
\end{figure}

In Fig. \ref{fig: env}, we plot the BER performance when employing different envelope-forming schemes. To ensure comparison fairness, all the schemes have the same spectral efficiency, and information bits carried by each subblock are allocated equally to two NOMA users. It should be noted that signals of the NU $(4,3,3)$ and the FU $(4,4,2)$ are based on the conventional OFDM waveform, while the OFDM-IM is applied to the remaining users. From Fig. \ref{fig: env} we observe that the OFDM-IM achieves a lower BER than the conventional OFDM in the high SNR regime. However, in the low SNR regime, the conclusion is the opposite. It can be explained that the inactive subcarriers in the OFDM-IM reduce the interference in NOMA, but the low SNR values result in the index error which degrades the error performance.

\vspace{-0.4 cm}
\subsection{Comparisons with IM-NOMA}
\vspace{-0.1 cm}
Here, let us illustrate the superiority of the proposed NOMA-IE over the IM-enable NOMA in existing works. 

\begin{figure} [t!]
	\centering
	\includegraphics[width = 3.4in] {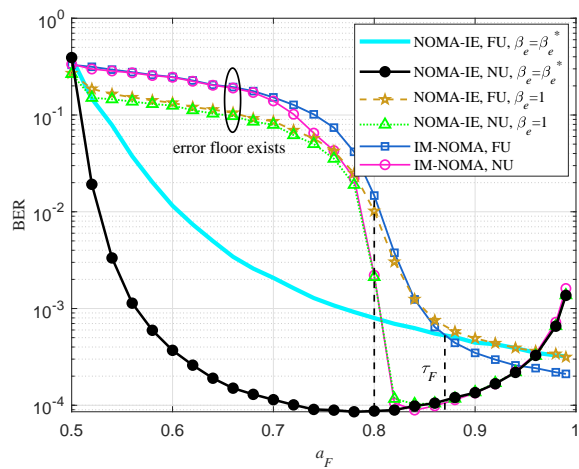}
	\caption{Error performance versus power allocation coefficient of the FU $a_F$ compared to IM-NOMA with SNR = 40 dB.}
	\label{fig: pacIM}
	\vspace{-0.2 cm}
\end{figure}

In Fig. \ref{fig: pacIM}, we compare the BER performance of NOMA-IE and IM-NOMA when $\alpha_F$ grows from 0.5 to 1. As seen from Fig. \ref{fig: pacIM}, when $a_F<\tau_F$ the NOMA-IE with the envelope detection coefficient ${\beta_e}^*$ considerably outperforms the IM-NOMA. When $a_F>\tau_F$, a slight performance gain can only be obtained at the FU of IM-NOMA. As mentioned previously, $a_F = 0.8$ is the bound deciding whether the error floor exists for IM-NOMA and NOMA-IE with the conventional detection method, i.e., $\beta_e = 1$. Since the definition of NOMA-IE ({\bf Definition \ref{def: IC-NOMA}}) guarantees that the constellations of the superimposed symbols are not overlapped under $a_F > a_N$, we are able to find a detection method to avoid the error floor. That is to say, the design of the NOMA-IE enlarges the feasible value range of $a_F$. Therefore, the NOMA-IE has the potential to provide more flexibility for different user requirements.
\begin{figure} [t!]
	\centering
	\includegraphics[width = 3.4in] {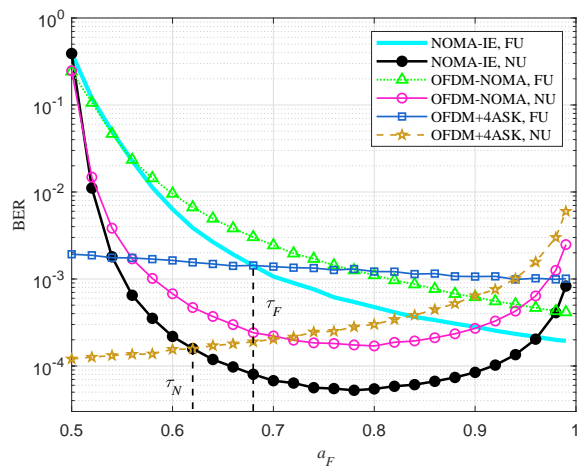}
	\caption{Error performance versus power allocation coefficient of the FU $a_F$ compared to conventional OFDM-based schemes with SNR = 40 dB.}
	\label{fig: paccon}
	\vspace{-0.2 cm}
\end{figure}

\vspace{-0.4 cm}
\subsection{Comparisons with Conventional OFDM-based Schemes}
\begin{figure*}[t!] 
	\centering
	\subfigure[]{\includegraphics[width=2.9in]{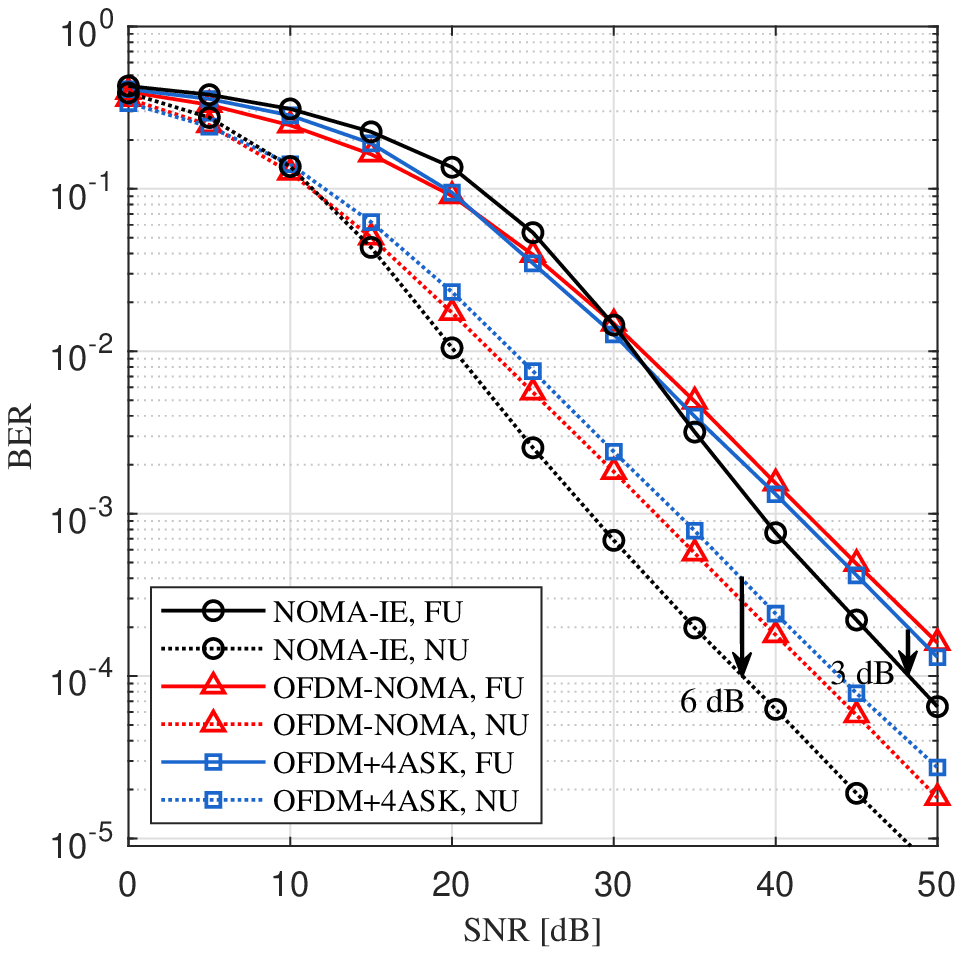}
		\label{c: fig_a}}
	\hfil
	\subfigure[]{\includegraphics[width=2.9in]{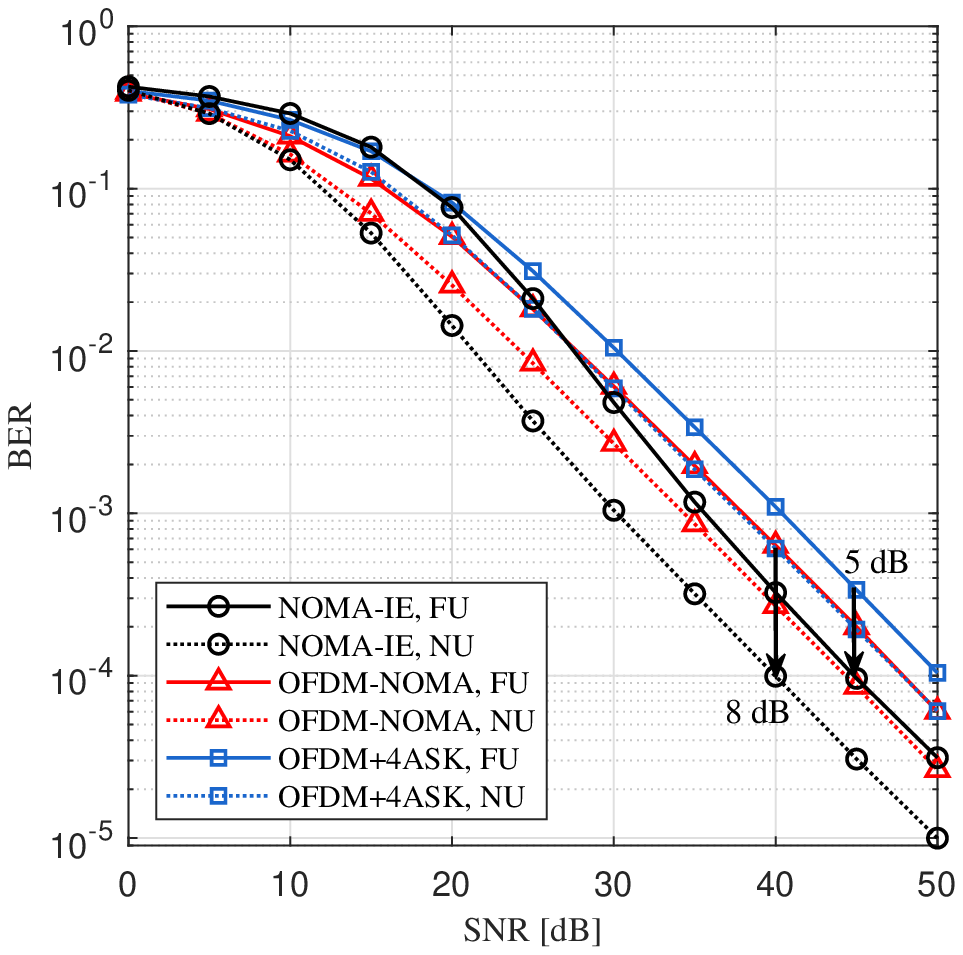}
		\label{c: fig_b}}
	\caption{Error performance versus SNR compared with conventional OFDM-based schemes: (a) $a_F = 0.75$, $a_N = 0.25$; (b) $a_F = 0.9$, $a_N = 0.1$.
	}
	\label{fig: comparisoncon}
\end{figure*}
In this subsection, we compare the proposed NOMA-IE scheme with conventional OFDM and OFDM-NOMA. The power reallocation policy is employed to guarantee the equal energy efficiency of the considered schemes.

In Fig. \ref{fig: paccon}, we present the BER performance of the three schemes versus the power allocation coefficient $\alpha_F$. It can be observed that with a BER value lower than $10^{-2}$, the proposed NOMA-IE always outperforms the OFDM-NOMA at the same spectral efficiency and energy efficiency. This observation validates the performance gain from the flexibility of the NOMA envelope. Compared with the conventional OFDM, the FU and the NU have better BER performance with $a_F > \tau_F$ and $ a_F > \tau_{N}$, respectively. Therefore, the NOMA-IE with appropriate power allocation coefficients is capable of achieving higher BER performance than the conventional OFDM-based schemes. 

Fig. \ref{fig: comparisoncon} presents the BER performance of the three schemes versus the SNR to show the exact gain from the NOMA-IE.
As shown in Fig. \ref{c: fig_a} and Fig. \ref{c: fig_b}, both the FU and the NU in NOMA-IE have approximately 4 dB better BER performance than OFDM-NOMA at the BER value of $10^{-4}$. The performance gain from NOMA-IE over OFDM of the FU is larger than that of the NU. It can be explained that the NOMA-IE is a mixture of OFDM-NOMA and OFDM and hence suffers less interference than the OFDM-NOMA. For the NU, up to 8 dB BER gain can be achieved at BER of $10^{-4}$ when $a_F = 0.9$. Moreover, it can be observed that obvious performance gains occur when the SNR is larger than 30 dB. Therefore, we draw the conclusion that the proposed NOMA-IE is effective in the high SNR regime.

\section{Conclusion}
\vspace{-0.3 cm}
In this paper, a novel NOMA-IE scheme based on the OFDM-IM framework has been proposed to explore a new degree of freedom for NOMA. It can be concluded that the envelope is beneficial to NOMA, especially in high SNR regimes. In particular, we have answered three questions listed in Section I to show the benefits of the signal envelope in NOMA. Firstly, the multi-subcarrier subblock can be expressed as the combination of multiple two-subcarrier subblocks in the NOMA-IE. Since the theoretical BER expressions of the two-subcarrier subblock are easily obtained, we are able to derive that of the multi-subcarrier subblock when considering the intra-cluster interference in NOMA. Secondly, the joint design of the envelope former and the envelope detection coefficient eliminates the error floor, conditioned that the higher power level is allocated to the NU than the FU. Lastly, under equal spectral efficiency and energy efficiency, we have shown that the NOMA-IE outperforms the OFDM in terms of error performance when the power allocation coefficient of the NU is larger than a specific threshold. We have also demonstrated that the design of NOMA-IE reduces the intra-cluster interference in NOMA, which improves the error performance of NOMA-IE, especially at high SNR.


\vspace{-0.5 cm}
\bibliographystyle{IEEEtran}
\bibliography{reference}

\begin{thebibliography}{10}
\providecommand{\url}[1]{#1}
\csname url@samestyle\endcsname
\providecommand{\newblock}{\relax}
\providecommand{\bibinfo}[2]{#2}
\providecommand{\BIBentrySTDinterwordspacing}{\spaceskip=0pt\relax}
\providecommand{\BIBentryALTinterwordstretchfactor}{4}
\providecommand{\BIBentryALTinterwordspacing}{\spaceskip=\fontdimen2\font plus
\BIBentryALTinterwordstretchfactor\fontdimen3\font minus
  \fontdimen4\font\relax}
\providecommand{\BIBforeignlanguage}[2]{{%
\expandafter\ifx\csname l@#1\endcsname\relax
\typeout{** WARNING: IEEEtran.bst: No hyphenation pattern has been}%
\typeout{** loaded for the language `#1'. Using the pattern for}%
\typeout{** the default language instead.}%
\else
\language=\csname l@#1\endcsname
\fi
#2}}
\providecommand{\BIBdecl}{\relax}
\BIBdecl

\bibitem{0conference}
Z.~{Xie}, W.~{Yi}, X.~{Wu}, Y.~{Liu}, and A.~{Nallanathan}, ``{NOMA} with
  informative envelope,'' in \emph{Proc. IEEE Int. Conf. Commun. (ICC)}, May
  2023, submitted.

\bibitem{6G}
W.~Saad, M.~Bennis, and M.~Chen, ``A vision of {6G} wireless systems:
  Applications, trends, technologies, and open research problems,''
  \emph{{IEEE} Netw.}, vol.~34, no.~3, pp. 134--142, 2020.

\bibitem{NOMA1}
L.~Dai, B.~Wang, Y.~Yuan, S.~Han, I.~Chih-lin, and Z.~Wang, ``Non-orthogonal
  multiple access for {5G}: solutions, challenges, opportunities, and future
  research trends,'' \emph{{IEEE} Commun. Mag.}, vol.~53, no.~9, pp. 74--81,
  2015.

\bibitem{NOMA2}
Z.~Ding, Z.~Yang, P.~Fan, and H.~V. Poor, ``On the performance of
  non-orthogonal multiple access in {5G} systems with randomly deployed
  users,'' \emph{{IEEE} Signal Process. Lett.}, vol.~21, no.~12, pp.
  1501--1505, 2014.

\bibitem{NOMA3}
Y.~Liu, Z.~Qin, M.~Elkashlan, Z.~Ding, A.~Nallanathan, and L.~Hanzo,
  ``Nonorthogonal multiple access for {5G} and beyond,'' \emph{Proc. {IEEE}},
  vol. 105, no.~12, pp. 2347--2381, 2017.

\bibitem{DingPairu}
Z.~Ding, P.~Fan, and H.~V. Poor, ``Impact of user pairing on {5G} nonorthogonal
  multiple-access downlink transmissions,'' \emph{{IEEE} Trans. Veh. Technol.},
  vol.~65, no.~8, pp. 6010--6023, 2016.

\bibitem{NOMAUAV}
N.~Zhao, X.~Pang, Z.~Li, Y.~Chen, F.~Li, Z.~Ding, and M.-S. Alouini, ``Joint
  trajectory and precoding optimization for {UAV}-assisted {NOMA} networks,''
  \emph{{IEEE} Trans. Commun.}, vol.~67, no.~5, pp. 3723--3735, 2019.

\bibitem{NOMAout2}
T.~N. Do, D.~B. da~Costa, T.~Q. Duong, and B.~An, ``Improving the performance
  of cell-edge users in noma systems using cooperative relaying,'' \emph{{IEEE}
  Trans. Commun.}, vol.~66, no.~5, pp. 1883--1901, 2018.

\bibitem{NOMAout3}
Z.~Xie, W.~Yi, X.~Wu, Y.~Liu, and A.~Nallanathan, ``{STAR-RIS} aided {NOMA} in
  multicell networks: A general analytical framework with gamma distributed
  channel modeling,'' \emph{{IEEE} Trans. Commun.}, vol.~70, no.~8, pp.
  5629--5644, 2022.

\bibitem{NOMAcapa}
W.~Yi, Y.~Liu, A.~Nallanathan, and M.~Elkashlan, ``Clustered millimeter-wave
  networks with non-orthogonal multiple access,'' \emph{{IEEE} Trans. Commun.},
  vol.~67, no.~6, pp. 4350--4364, 2019.

\bibitem{IMOFDMBasar}
E.~Ba\c{s}ar, {\"U}.~Ayg\"{o}l\"{u}, E.~Panay{\i}rc{\i}, and H.~V. Poor,
  ``Orthogonal frequency division multiplexing with index modulation,''
  \emph{{IEEE} Trans. Signal Process.}, vol.~61, no.~22, pp. 5536--5549, 2013.

\bibitem{IM1}
E.~Basar, ``Index modulation techniques for {5G} wireless networks,''
  \emph{{IEEE} Commun. Mag.}, vol.~54, no.~7, pp. 168--175, 2016.

\bibitem{IM2}
M.~Wen, X.~Cheng, M.~Ma, B.~Jiao, and H.~V. Poor, ``On the achievable rate of
  {OFDM} with index modulation,'' \emph{{IEEE} Trans. Signal Process.},
  vol.~64, no.~8, pp. 1919--1932, 2016.

\bibitem{IMSNR}
S.~Dang, J.~P. Coon, and G.~Chen, ``Adaptive {OFDM} with index modulation for
  two-hop relay-assisted networks,'' \emph{{IEEE} Trans. Wireless Commun.},
  vol.~17, no.~3, pp. 1923--1936, 2018.

\bibitem{IMsurvey}
T.~Mao, Q.~Wang, Z.~Wang, and S.~Chen, ``Novel index modulation techniques: A
  survey,'' \emph{{IEEE} Commun. Surv. Tuts.}, vol.~21, no.~1, pp. 315--348,
  2019.

\bibitem{BasarSurv}
E.~Basar, M.~Wen, R.~Mesleh, M.~Di~Renzo, Y.~Xiao, and H.~Haas, ``Index
  modulation techniques for next-generation wireless networks,'' \emph{IEEE
  Access}, vol.~5, pp. 16\,693--16\,746, 2017.

\bibitem{SubIM}
R.~Abu-alhiga and H.~Haas, ``Subcarrier-index modulation {OFDM},'' in
  \emph{Proc. IEEE Int. Symp. Pers. Indoor Mobile Radio Commun. (PIMRC)}, 2009,
  pp. 177--181.

\bibitem{IMNOMA}
E.~Arslan, A.~T. Dogukan, and E.~Basar, ``Index modulation-based flexible
  non-orthogonal multiple access,'' \emph{{IEEE} Wireless Commun. Lett.},
  vol.~9, no.~11, pp. 1942--1946, 2020.

\bibitem{IMNOMAopen}
A.~Almohamad, M.~O. Hasna, S.~Althunibat, and K.~Qaraqe, ``A novel downlink
  {IM-NOMA} scheme,'' \emph{{IEEE} Open J. Commun. Soc.}, vol.~2, pp. 235--244,
  2021.

\bibitem{IETBER}
F.~Kara and H.~Kaya, ``{BER} performances of downlink and uplink {NOMA} in the
  presence of {SIC} errors over fading channels,'' \emph{{IET} Commun.},
  vol.~12, no.~15, pp. 1834--1844, 2018.

\bibitem{SER}
Q.~He, Y.~Hu, and A.~Schmeink, ``Closed-form symbol error rate expressions for
  non-orthogonal multiple access systems,'' \emph{{IEEE} Trans. Veh. Technol.},
  vol.~68, no.~7, pp. 6775--6789, 2019.

\bibitem{BERBPSK}
M.~Aldababsa, C.~Göztepe, G.~K. Kurt, and O.~Kucur, ``Bit error rate for
  {NOMA} network,'' \emph{{IEEE} Commun. Lett.}, vol.~24, no.~6, pp.
  1188--1191, 2020.

\bibitem{BERArbi}
H.~Yahya, E.~Alsusa, and A.~Al-Dweik, ``Exact {BER} analysis of {NOMA} with
  arbitrary number of users and modulation orders,'' \emph{{IEEE} Trans.
  Commun.}, vol.~69, no.~9, pp. 6330--6344, 2021.

\bibitem{HybridIM}
A.~Tusha, S.~Doğan, and H.~Arslan, ``A hybrid downlink {NOMA} with {OFDM} and
  {OFDM-IM} for beyond {5G} wireless networks,'' \emph{{IEEE} Signal Process.
  Lett.}, vol.~27, pp. 491--495, 2020.

\bibitem{IMNOMAup}
S.~Doğan, A.~Tusha, and H.~Arslan, ``{NOMA} with index modulation for uplink
  {URLLC} through grant-free access,'' \emph{{IEEE} J. Sel. Topics Signal
  Process.}, vol.~13, no.~6, pp. 1249--1257, 2019.

\bibitem{efloor}
M.~B. Shahab, S.~J. Johnson, M.~Shirvanimoghaddam, M.~Chafii, E.~Basar, and
  M.~Dohler, ``Index modulation aided uplink {NOMA} for massive machine type
  communications,'' \emph{{IEEE} Wireless Commun. Lett.}, vol.~9, no.~12, pp.
  2159--2162, 2020.

\bibitem{OFDMNOMA}
Y.~Xie, K.~C. Teh, and A.~C. Kot, ``Deep learning-based joint detection for
  {OFDM-NOMA} scheme,'' \emph{{IEEE} Commun. Lett.}, vol.~25, no.~8, pp.
  2609--2613, 2021.

\bibitem{IMpower}
M.~Irfan and S.~Aïssa, ``Generalization of index-modulation: Breaking the
  conventional limits on spectral and energy efficiencies,'' \emph{{IEEE}
  Trans. Wireless Commun.}, vol.~20, no.~6, pp. 3911--3924, 2021.

\bibitem{PEP}
M.~K. Simon and M.-S. Alouini, \emph{Digital Communication over Fading
  Channels}.\hskip 1em plus 0.5em minus 0.4em\relax New York, USA: Wiley, 2000.

\bibitem{Qfappr}
M.~Chiani, D.~Dardari, and M.~Simon, ``New exponential bounds and
  approximations for the computation of error probability in fading channels,''
  \emph{{IEEE} Trans. Wireless Commun.}, vol.~2, no.~4, pp. 840--845, 2003.

\end{thebibliography}


\end{document}